# Unexpected novel *Merbecovirus* discoveries in agricultural sequencing datasets from Wuhan, China

Daoyu Zhang[1], Adrian Jones[2], Yuri Deigin[3*], Karl Sirotkin[4], Alejandro Sousa[5,6]

[1] Independent genetics researcher
[2] Independent bioinformatics researcher
[3] Youthereum Genetics Inc., Toronto, Ontario, Canada
[4] Karl Sirotkin LLC, Lake Mary, FL 32746, USA
https://orcid.org/0000-0002-9685-0338
[5] Regional Hospital of Monforte, Lugo, Spain
[6] University of Santiago de Compostela, Spain

*Correspondence to: ydeigin@protonmail.com

## Abstract

In this study we document the unexpected discovery of multiple coronaviruses and a BSL-3 pathogen in agricultural cotton and rice sequencing datasets. In particular, we have identified a novel HKU5-related *Merbecovirus* in a cotton dataset sequenced by the Huazhong Agricultural University in 2017. We have also found an infectious clone sequence containing a novel HKU4-related *Merbecovirus* related to MERS coronavirus in a rice dataset sequenced by the Huazhong Agricultural University in early 2020. Another HKU5-related *Merbecovirus*, as well as Japanese encephalitis virus, were identified in a cotton dataset sequenced by the Huazhong Agricultural University in 2018. An HKU3-related *Betacoronavirus* was found in a *Mus musculus* sequencing dataset from the Wuhan Institute of Virology in 2017. Finally, a SARS-WIV1-like *Betacoronavirus* was found in a rice dataset sequenced by the Fujian Agriculture and Forestry University in 2017. Using the contaminating reads we have extracted from the above datasets, we were able to assemble complete genomes of two novel coronaviruses which we disclose herein. In light of our findings, we raise concerns about biosafety protocol breaches, as indicated by our discovery of multiple dangerous human pathogens in agricultural sequencing laboratories in Wuhan and Fouzou City, China.

## Introduction

The highly complex and often poorly curated SRA data submitted by institutions to NCBI may sometimes offer clues to the wider context of various laboratory activities taking place in the labs from which the SRA sequencing data originates – through either cross-contamination between samples or index hopping within multiplexed runs (Ballenghien et al. 2017). Therefore, identification and characterization of cross-contaminating reads from such data could provide useful insight about genetic sequences that might have been withheld or remain unpublished by these institutions.

Considering the recent scrutiny of the Wuhan Institute of Virology (WIV) and its potential connection to the outbreak of SARS-CoV-2 (Sirotkin and Sirotkin, 2020; Segreto and Deigin, 2020; Segreto et al. 2020; Joint WHO-China Study, 2021) we believe that a thorough metagenomic analysis of SRA reads from the WIV and other laboratories in Wuhan, China could potentially help in research into the origins of SARS-CoV-2.

Here we carry out an exploration of several NCBI SRA datasets containing the results of sequencing work conducted mostly in Wuhan over the past four years. Our analysis has yielded several viruses anomalous to the sequencing of rice and cotton datasets from Huazhong Agricultural University (HZAU) and Fujian Agriculture and Forestry University (FAFU), as well as a *Mus musculus* sequencing project from WIV.

In particular, we report the unexpected identification and full-length sequence assembly of a novel *Merbecovirus* (a *Betacoronavirus* subgenus) of the HKU5 family in cotton sequencing BioProject PRJNA396502 in Hi-C sequencing data of *Gossypium barbadense* (NCBI SRA accession SRR5885860), deposited by the Huazhong Agricultural University (HZAU) in 2017. In addition, we report the identification of an additional *Merbecovirus* sequence from raw sequence reads of *Gossypium Hirsutum* in cotton sequencing BioProject PRJNA380842, deposited by HZAU (NCBI SRA accession SRR7896912) nearly a year before the official submission of the genetic sequence (accession MN611520.1) and its sequence read archive (SRA) to NCBI GenBank by the WIV.

Furthermore, we have undertaken *de novo* assembly of a dataset from an *Oryza sativa Japonica* sequencing BioProject (PRJNA602160) and isolated a putative human infectious clone sequence containing a novel HKU4-r CoV. The clone was found to contain a hepatitis delta virus (HDV) ribozyme at its 3'-end in a format very similar to that of a Nipah virus infectious clone identified by Zhang (2021) from BioProject PRJNA605983 containing samples from five patients infected with COVID-19 at the early stage of the Wuhan seafood market outbreak amplification event (Zhou et al. 2020).

We also document the identification of biosafety level (BSL)-3 pathogen Japanese encephalitis virus (JEV) in cotton sequencing BioProject PRJNA380842, as well as a human infectious SARS-WIV1 CoV (Ge et al. 2013, Menachery et al. 2015) in rice sequencing BioProject PRJNA358135, and a HKU3-related CoV in *Mus musculus* sequencing BioProject PRJNA393936.

Considering that bat CoVs HKU4 and HKU5 are Middle East respiratory syndrome-related coronaviruses (MERS-r CoV) (Wang et al. 2014) and that MERS-CoV is a BSL-3 level pathogen (Algaissi et al. 2020), MERS-r CoV related research should be a highly regulated process, conducted within high-containment biosafety laboratories. The presence of contaminating reads of potentially human pathogenic MERS-r CoVs within agricultural research datasets, which do not have the same high biosafety standards or protocols as BSL3 laboratories (Klein, 2012; Heckert and Kozlovac, 2014), presents evidence of potential significant public health hazards associated with biosafety laboratories in Wuhan, China. Furthermore, the discovery of the unexpected *Merbecovirus* sequences, which were actively researched by WIV (Yang et al. 205, Luo et al. 2018), in HZAU datasets indicates potential cross-contamination between WIV and HZAU, both between samples and within multiplexed runs.

Moreover, the fact that the above contamination was not identified by the original project authors potentially points to even more widespread contamination between these laboratories. The discovery of these unpublished viruses sequenced by WIV and HZAU, also brings into question the possible existence of other unpublished coronavirus strains at WIV and other Wuhan laboratories.

Also, the fact that the two novel coronaviruses documented here were sequenced (and at least one actively researched) but have not yet been published points to a pattern of behavior of cloning and research of coronaviruses by the WIV without their disclosure to the wider international community. Furthermore, the identification of these sequences as contaminants in low-risk agricultural datasets indicates that these samples may have been sequenced without notifying the sequencing institution of their potential pathogenicity.

In summary, the discoveries we present herein are indicative of biocontainment security failures within laboratories in Wuhan that may have relevance in the search for SARS-CoV-2 origins.

## Materials & Methods

### Datasets

Multiple datasets deposited by the HZAU and the WIV to the NCBI SRA database (Agarwala et al. 2018) were sampled and checked for reads associated with human-pathogenic viruses using the public NCBI SRA Taxonomy Analysis Tool (STAT) (Katz et al., 2021), with a focus on datasets of an agricultural orientation in experimental design and sample origin. Datasets found to contain reads associated with human and vertebrate viral families were subjected to further analysis. Fastv (Chen et al. 2020) was run for each SRA against the Opengene vial genome kmer collection ‘microbial.kc.fasta.gz’ (https://github.com/OpenGene/UniqueKMER). SRA format datasets from NCBI were converted to either paired or single fastq files (depending on layout) using sratoolkit version 2.10.9 (https://trace.ncbi.nlm.nih.gov/Traces/sra/sra.cgi?view=software).

### PRJNA396502: (including novel HKU5-related sequence) sequence identification and assembly

NCBI STAT analysis identification of vertebrate viral families and in particular human pathogenic viral families was conducted for each dataset in this BioProject.

Where viruses were identified, an NCBI BLAST (Johnson et al. 2008) search using a reference sequence of the most specific STAT taxonomy was conducted against the dataset in order to obtain the presence and coverage of virus-related reads within the dataset. Whenever a significant hit was found, the dataset was downloaded and *de novo* assembled using EGassembler (Masoudi-Nejad et al., 2006) and MEGAHIT (Li et al., 2015) to obtain contiguous sequences of potential viral origin. After verification of the contiguous sequences, contigs from the same dataset were aligned together using HKU5 as a reference (accession NC_009020.1) and a genome-length consensus sequence of a HKU5-related *Merbecovirus* was successfully assembled from reads obtained from SRR5885860.

For identifying potential host sequences from contaminating reads, sequences from mitochondrial DNA were retrieved from the datasets using NCBI BLAST (Johnson et al. 2008) search and assembled using the PRABI-Doua CAP3 sequence assembly program (http://doua.prabi.fr/software/cap3) (Huang and Madan, 1999). The resulting assembled contigs and unassembled sequences were then identified using NCBI BLAST (Johnson et al. 2008) search against the NCBI nr/nt database (Agarwala et al. 2018).

**SRR5885860 (BioProject PRJNA396502): Identifying ORFs from the consensus sequence of the novel HKU5-related CoV**

After *de novo* assembly of SRR5885860 and consensus sequence extraction, Addgene sequence analyzer (https://www.addgene.org/analyze-sequence/) was used to identify Open Reading Frames (ORFs) from the consensus sequence, which revealed a typical *Merbecovirus* genomic configuration. After identification of the *Merbecovirus* genome, the ORFs within the sequence were sequentially searched following each instance of the *Merbecovirus* transcription-regulating sequence (TRS), “AACGAAC”, using the APE sequence analysis and annotation tool version 2.0.59t (https://jorgensen.biology.utah.edu/wayned/ape/). All open reading frames of the *Merbecovirus* genome were identified and the genome was successfully annotated without problems, although the N TRS has been found to be “AACGAAT”, which likely resulted in the reduction of the N and ORF8b mRNA in this species.

**SRR5885860 (BioProject PRJNA396502): Novel HKU5-related CoV phylogenetic analysis**

For RdRp and spike protein analysis, BlastAlign (Belshaw and Katzourakis, 2005) with other partial RdRps/spike proteins was used to identify the respective range on the query sequence. The RdRp and the spike protein of the newly identified *Merbecovirus* was subjected to phylogenetic analysis by performing a NCBI BLAST (Johnson et al. 2008) search against the NCBI nr/nt dataset. The BLAST results were used to construct a phylogenetic tree using the NCBI BLAST distance tree of results tool with default settings.

Alignment of the spike glycoprotein to other *Merbecoviruses* was generated using Multalin (Corpet, 1988).

**SRR5885860, SRR5885851,SRR5885852, SRR5885879 (BioProject PRJNA396502): Alignment to reference genomes.**

Dataset SRR5885860 was aligned to MN6115201.1, while datasets SRR5885851, SRR5885852, SRR5885879 were pooled then aligned to each of NC_001407.1, NC_029853.1, NC_043404.1 using bowtie2 version 2.4.2 (Langmead and Salzberg,

2012). GATK version 4.1.9.0 (Van der Auwera & O'Connor, 2020) was used to sort then mark duplicates. Samtools version 1.11 (Li et al. 2009) was used to index the marked .bam files for viewing in IGV version 2.8.13 (Thorvaldsdóttir et al. 2013). Consensus sequences for the three pooled alignments to reference sequences were generated using IGV.

**PRJNA602160: (including HKU-4 related sequence) data and assembly**
Analysis of each SRA experiment in *Oryza sativa Japonica* group sequencing project accession PRJNA602160 was conducted using the NCBI STAT phylogenetic analysis tool accessed via the 'Analysis' tab for each SRA run. BtTp-BetaCoV/GX2012 (accession: KJ473822.1) was identified as a genome with significant matches to SRA run accessions: SRR10915168, SRR10915173 and SRR10915174. Run SRR10915167 exhibited matches to *Tylonycteris* bat coronavirus HKU4 (accession: NC_009019.1) and MERS (accession: NC_019843.3) Merbecoviruses.

The four sequencing datasets were downloaded and each was assembled individually using MEGAHIT v1.2.9 (Li et al. 2015) with default settings. The resulting contig sequences were then aligned using Nucleotide BLAST against BtTp-BetaCoV/GX2012 (accession: KJ473822.1). A complete sequence 32725 nt in length from accession SRR10915173 was identified as being 98.38% similar to the closest related sequence on NCBI, BtTp-BetaCoV/GX2012 (accession: KJ473822.1). An attempt to search for the natural host of HKU4-related Coronaviruses, the *Tylonycteris pachypus* bat, were performed on this dataset, however, no sequences could be found to identify it as the species.

To confirm the MEGAHIT assembly we also undertook *de novo* assembly of SRR10915173 using coronaSPAdes v3.15.2 (Meleshko et al. 2021) using default settings, and SPAdes v3.15.2 (Prjibelski et al. 2020) using the '--careful' parameter. We then used bowtie2 version 2.4.2 (Langmead and Salzberg, 2012) to align the final contigs from both applications to the 32725 nt long contig 'k141_11929' generated by the MEGAHIT assembly, which contained the novel HKU4-r CoV. For coronaSPAdes, two contigs fully covered the sequence, while for SPAdes a single contig covered the entire sequence.

Raw read depth coverage analysis over the MEGAHIT assembly contig K141_11929 (containing the novel HKU4-r CoV) was also conducted (Supp. Fig. 22).

**SRR10915173 (BioProject PRJNA602160): Identification of the HKU4-related sequence as an infectious clone**
As the contig sequence K141_11929 obtained via *de novo* assembly of SRR10915173 was found to be longer than the genome size of Merbecoviruses HKU4 (30247nt), we performed a NCBI BLAST (Johnson et al. 2008) analysis of the sequences flanking the HKU4-related genome on this contig, which revealed homology to many expression and cloning vector sequences that were directly fused to the 5'-end and the 3'-end of the coronavirus genome. A BLAST search of the 5'-end and 3'-end of the coronavirus genome was performed, which verified the presence of reads covering the vector-virus junctions on both the 5'-end and 3'-end of the genome.
Sequence analysis was then performed using the Addgene sequence analyzer (https://www.addgene.org/analyze-sequence/), which revealed a human cytomegalovirus (CMV) promoter before the 5'-end of the HKU4-related CoV genome and a bovine growth hormone polyadenylation (bgH polyA) signal after the 3'-end of the HKU4-related CoV genome, confirming the sequence origin as a construct intended to be infectious. The complete genome of the HKU4-related coronavirus was manually annotated to indicate all open reading frames (ORFs).

**SRR10915173 (BioProject PRJNA602160): Characterization of the cloned HKU4-related sequence**

Sequence characterization using NCBI nucleotide BLAST (BLASTn) (Johnson et al. 2008) and RDP5 (Martin et al. 2020) were performed, which revealed the HKU4-related CoV sequence as only 98.38% similar to the closest sequence of HKU4, KJ473822.1.

**SRR10915173 (BioProject PRJNA602160): Modeling the Receptor Binding Domain of the cloned HKU4-related CoV's Spike**

Structure of the Receptor Binding Domain (RBD) for the novel HKU4-r CoV was modeled using SWISS-MODEL webserver (https://swissmodel.expasy.org/) (Waterhouse et al., 2018) and aligned to PDB id: 4QZV using PyMol webserver Version 2.4 (https://pymol.org/2/) (The PyMOL Molecular Graphics System). The binding free energy of the complex was calculated using PRODIGY webserver (https://bianca.science.uu.nl/prodigy/) (Xue et al., 2016) and compared against that of the canonical complex PDB id: 4QZV.

**PRJNA358135: SARS-WIV1 CoV assembly**

Eight paired end SRA datasets (SRR5127175, SRR5127197, SRR5127192, SRR5127191, SRR5127189, SRR5127187, SRR5127186, SRR5127184) were pooled from project PRJNA358135 and *de novo* assembled using MEGAHIT with default settings. The raw reads were then aligned using the bat SARS-like coronavirus WIV1 complete genome (KF367457.1) full sequence as a reference using both bwa-mem version 0.7.17 with default settings and bowtie2 version 2.4.2 (Langmead and Salzberg, 2012) using the options "--local --very-sensitive-local". bwa version 0.7.17 was used to align the raw reads from the 8 pooled datasets to accession KF367457.1. GATK version 4.1.9.0 (Van der Auwera & O'Connor, 2020) was used to sort then mark duplicates. Samtools version 1.11 (Li et al. 2009) was used to index the marked .bam file for viewing in IGV version 2.8.13 (Thorvaldsdóttir et al. 2013). Consensus sequences for the two alignment methods generated using IGV.

**PRJNA380842: (Including HKU5-related CoV and JEV)**

NCBI STAT analysis and fastv (Chen et al. 2020) were run on each of SRR7896912, SRR7896916, SRR7896926 and SRR7896936. From NCBI STAT results we have identified the presence of MN611520.1 sequences in SRR7896912. We ran NCBI BLAST of SRX4734343 using MN611520.1 as a query.

The mitochondrion sequence for *Pipistrellus abramus*, the host of HKU5, occurs in RNA sequencing study SRR11085738 for *Pipistrellus abramus* bat coronavirus HKU5-related isolate BY140568 (accession MN611520.1). We attempted to find this mitochondrion reference genome sequence (NC_005436.1) in SRR7896912 by running a BLAST query of NC_005436.1 against the SRX4734343 experiment (Supp. Fig. 11). The hits were downloaded then queried in BLAST against the nt database. The following alignments were generated using the workflow below: SRR7896912 to MN611520.1; SRR7896916 to MN639770.1; SRR7896926 to NC_033138.1. Alignment was conducted using bowtie2 version 2.4.2 (Langmead and

Salzberg, 2012) with default parameters. GATK version 4.1.9.0 (Van der Auwera & O'Connor, 2020) was used to sort then mark duplicates. Samtools version 1.11 (Li et al. 2009) was used to index the marked .bam files for viewing in IGV version 2.8.13 (Thorvaldsdóttir et al. 2013).

**PRJNA393936: novel HKU3-r CoV discovery**

NCBI STAT was run on the four *Mus musculus* splenocyte sequencing runs in BioProject PRJNA393936. HKU4 and HKU3 sequences were identified in STAT results. Run accession SRR5819071 was then subjected to BLASTn analysis against HKU3-related CoV KF294457.1 using default settings. The 100 basalt results were downloaded as a complete sequence, then this complete sequence was queried in BLAST against the nt database.

## Results

**HKU5-related CoV discovery in BioProject PRJNA396502**

We have identified MERS-related coronavirus (MERS-r CoV) sequencing reads from SRA run SRR5885860 in the cotton 3D genome sequencing BioProject PRJNA396502 during a routine exploration of the NCBI SRA dataset for genome sequencing in Wuhan laboratories in 2017-2020 (Fig 1).

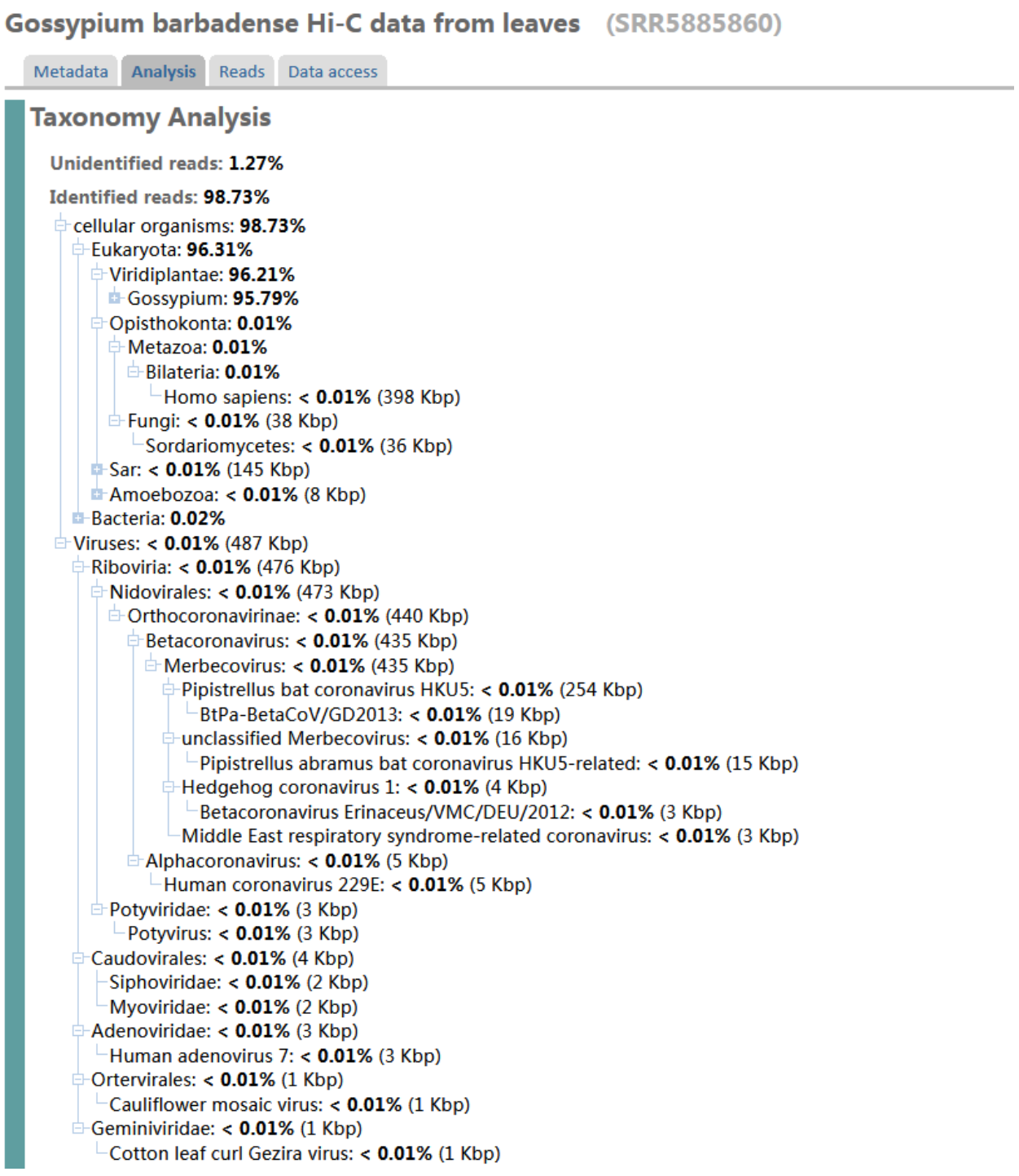

Fig. 1: NCBI SRA Taxonomy Analysis Tool (STAT) analysis results for SRR5885860.

Using the Coronavirus genome BtPa-BetaCoV/GD2013 (GenBank accession: KJ473820.1), we performed a nucleotide BLAST (Johnson et al. 2008) search on SRA sequencing run SRR5885860. This revealed more than 5000 sequences from a novel *Merbecovirus* closely related to HKU5 (Supp. Fig. 1).

We downloaded the sequence and used EGassembler (Masoudi-Nejad et al. 2006) to assemble the sequence, which revealed 13 contiguous sequences that were then mapped to the HKU5 *Merbecovirus* genome (Fig. 2).

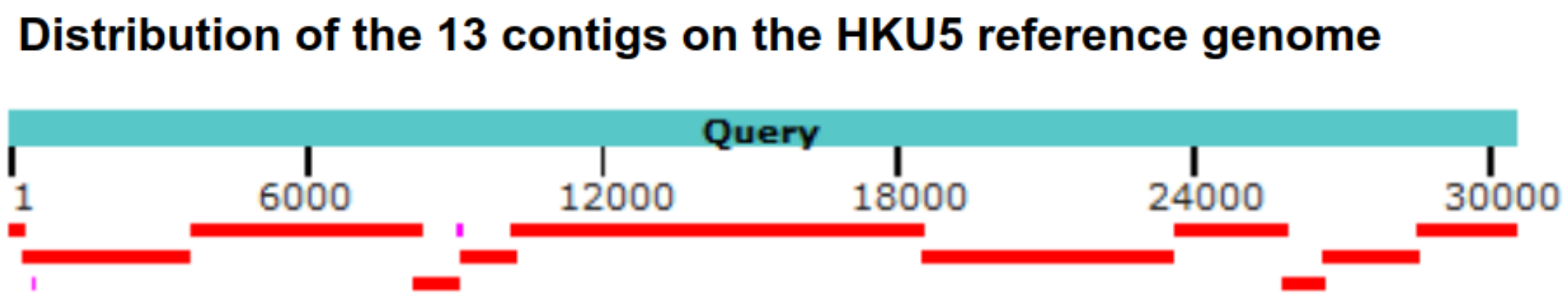

Fig. 2. Mapping of the 13 contigs onto the HKU5 genome.

A consensus sequence was assembled from the 13 contigs obtained from the *de novo* assembly result. We then analyzed the consensus sequence using Addgene sequence analyzer (https://www.addgene.org/analyze-sequence/) and the genome was found to have a typical *Merbecovirus* gene arrangement (Fig. 3).

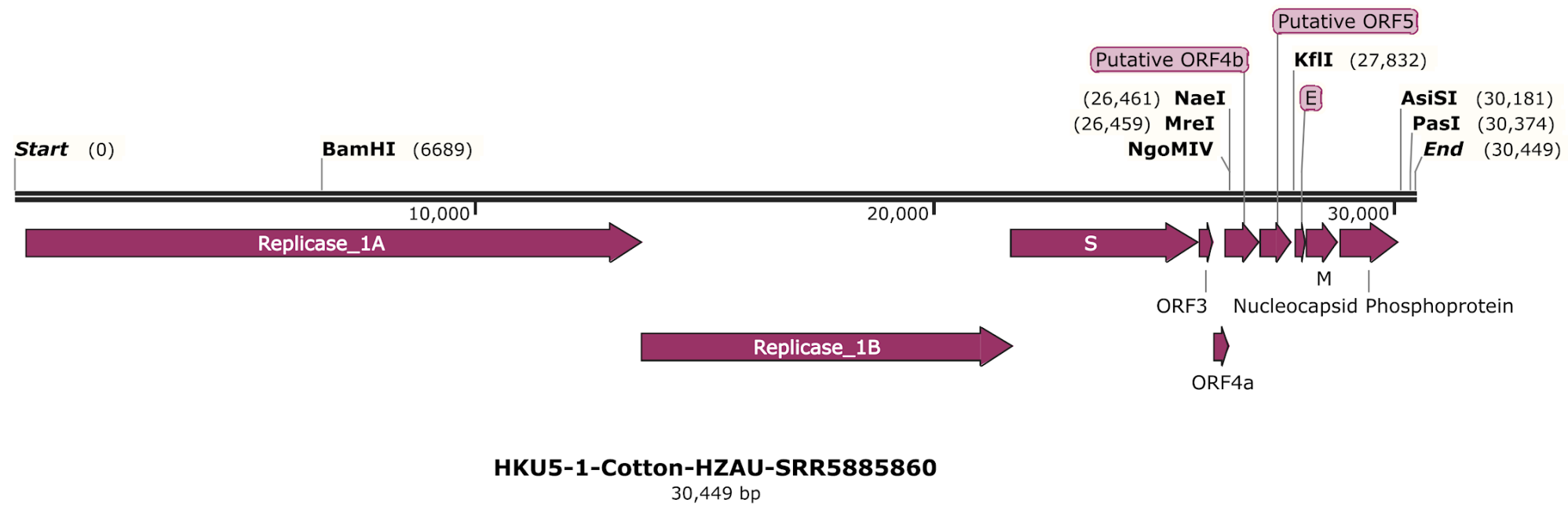

Fig. 3. Architectural mapping of the consensus sequence of the 13 Contigs obtained from SRR5885860 analyzed in Addgene sequence analyzer (https://www.addgene.org/analyze-sequence/), here displayed using SnapGene Viewer (SnapGene® software).

In order to validate the finding, MEGAHIT (Li et al., 2015) was used for independent *de novo* assembly of reads from SRR5885860, which revealed that a single 30440 nt contig contained a complete genome of a HKU5-related Merbecovirus that is nine nucleotides shorter at the 5'-end and seven nucleotides different from that obtained using EGassembler. The minor difference within the sequence potentially indicates the presence of Single Nucleotide Polymorphism (SNP) within this sample.

A BLAST analysis of the EGassembler-generated sequence against the NCBI nr/nt database revealed the sequence to be a divergent member of the HKU5 family of Merbecoviruses (Supp. Fig. 2). The sequence was found to be only 91% similar to other known full-length coronavirus sequences available at NCBI. Phylogenetic analysis was conducted which placed the novel Merbecovirus sequence in the HKU5 clade (Fig. 4, and Supp. Data: Supplementary_Information_4_HKU5-1-Cotton-HZAU-SRR5885860_Blast_Tree_View.pdf).

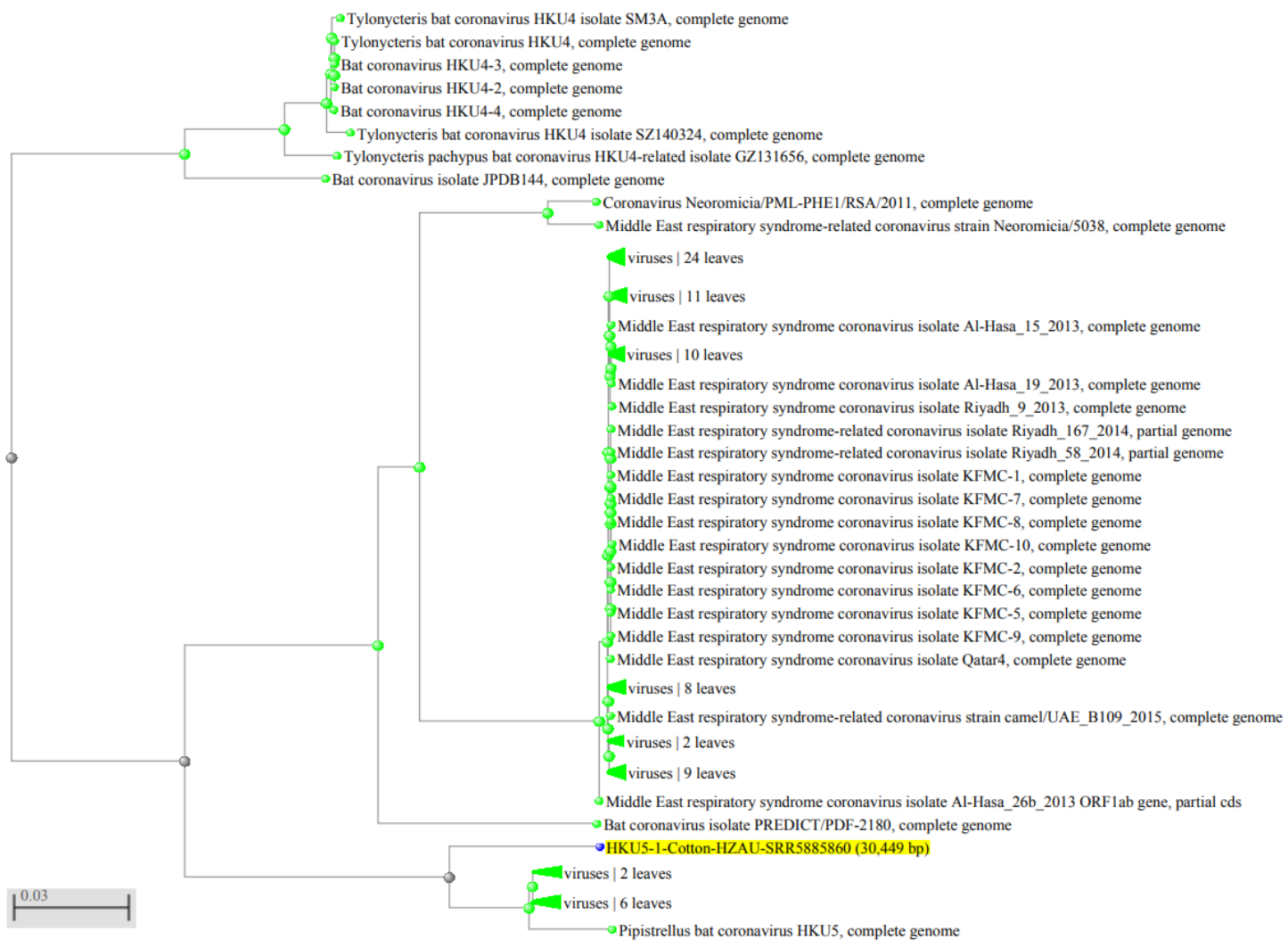

Fig. 4: Phylogenetic analysis places the novel Merbecovirus sequence in the HKU5 clade. The novel HKU5-CoV is highlighted in yellow.

BLAST analysis was then run on the gene coding region for RNA-dependent RNA polymerase (RdRp) (nt positions 15522-15908) against the available partial RdRp sequences at GenBank (Supp. Fig 3). An RdRp-based phylogenetic tree of the newly discovered Merbecovirus against available partial RdRp sequences on GenBank was then generated (Fig. 5).

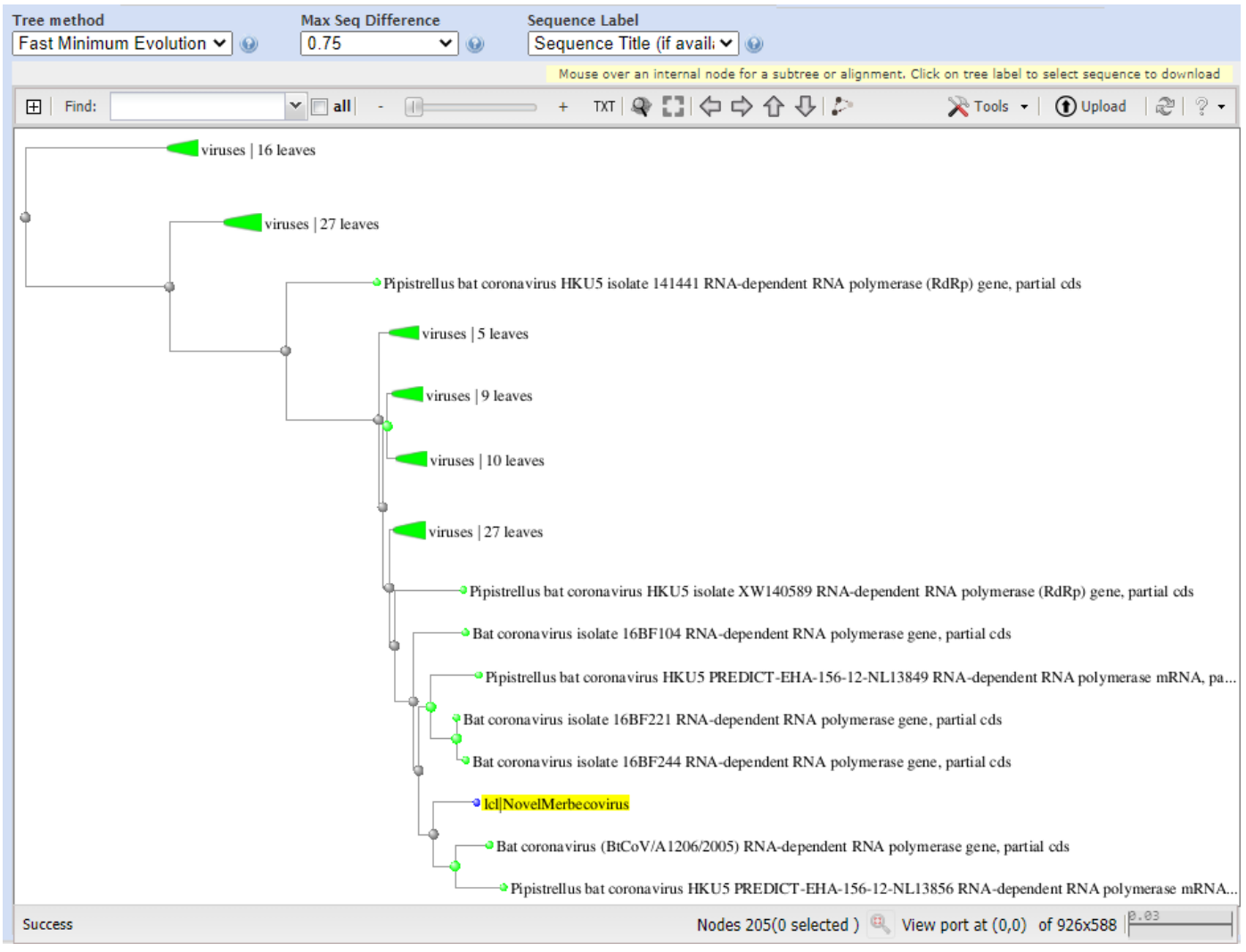

Fig. 5. RdRp based phylogenetic tree analysis of the newly discovered *Merbecovirus* against available partial RdRp sequences on GenBank.

To characterize the spike glycoprotein, BLAST analysis of the spike gene (nt positions 21710-25789) of the novel HKU5-related *Merbecovirus* against the BLAST nt database was conducted (Supp. Fig. 4). The coding sequence has the highest identity to that of *Pipistrellus* bat coronavirus HKU5, but was still significantly different from known sequences available at GenBank (Fig. 6).

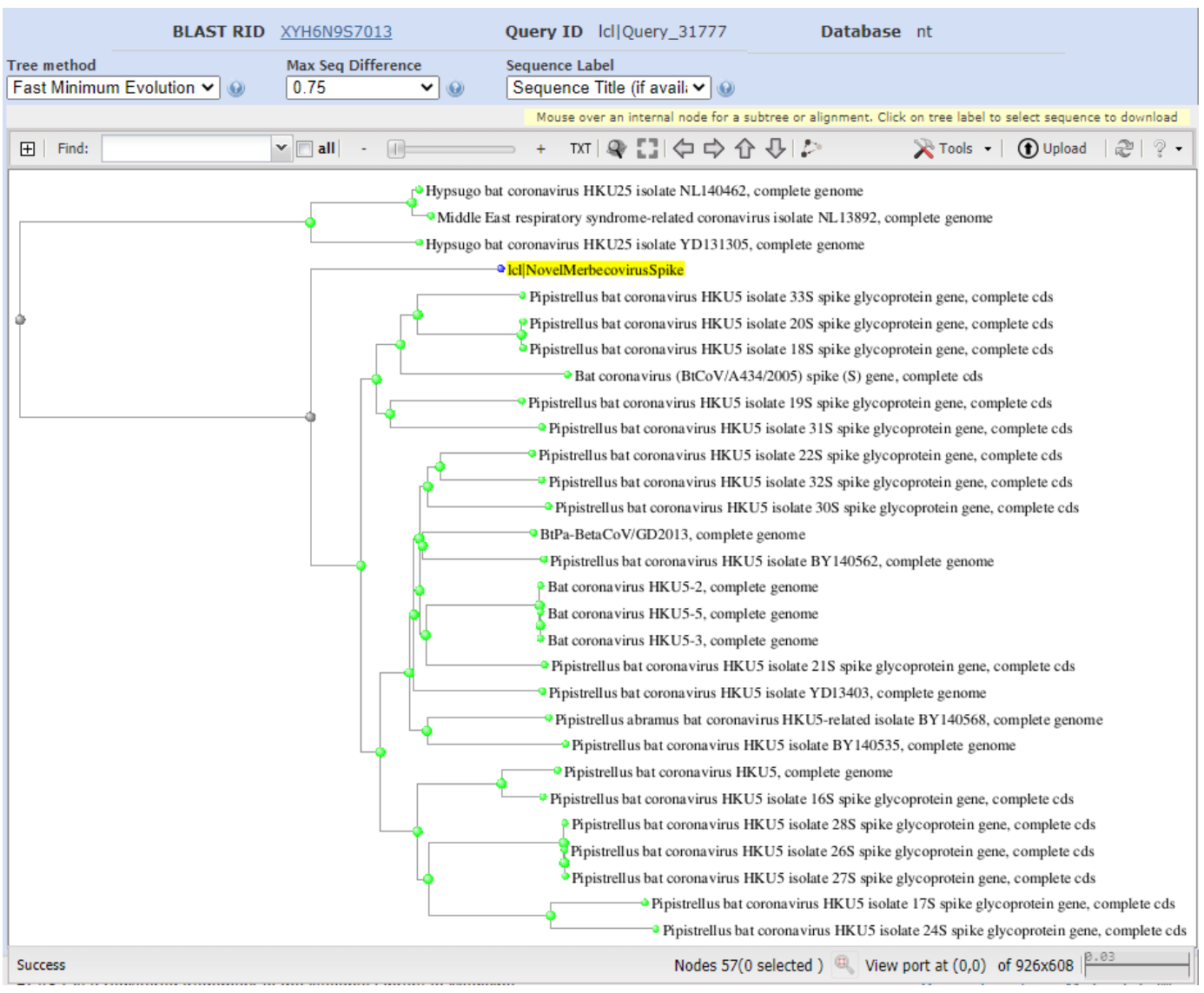

Fig. 6. Spike based phylogenetic tree of the novel HKU5-related *Merbecovirus* (nt positions 21710-25789) plotted against the closest spike sequences on GenBank.

The spike of this novel *Merbecovirus* shares only a 88.67% identity with the closest published spike in GenBank. As the 10 closest viruses by the RdRp region did not have published spike sequences, we cannot determine the spike similarity between this novel *Merbecovirus* and these viral strains. Closely related viruses for which both RdRp and spike sequences were published were found to have only a weak correlation between RdRp and spike identity (Fig. 7). We also cannot rule out a potential recombination event between these viral strains and the novel *Merbecovirus* identified in our study.

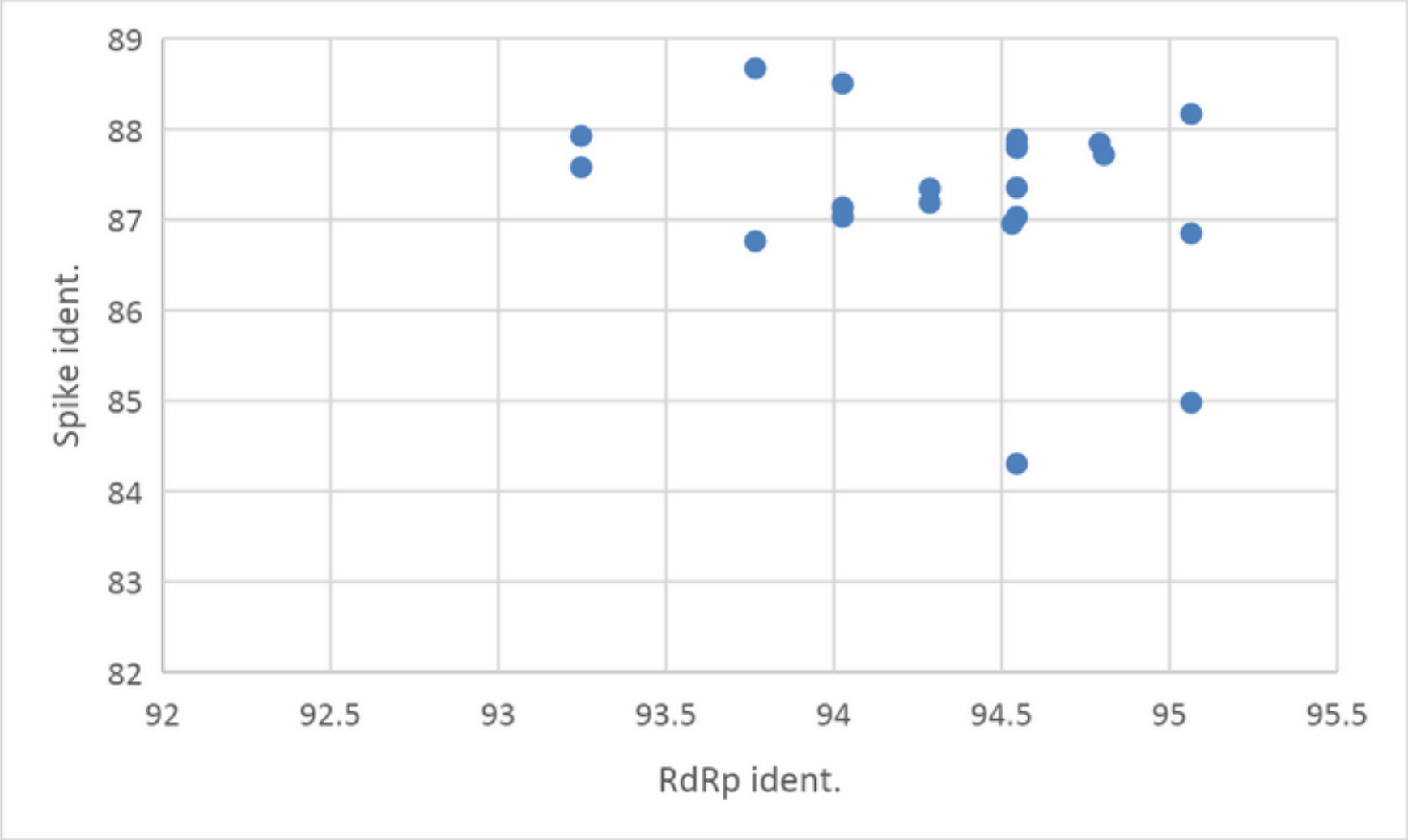

Fig. 7. RdRp and Spike identity plot of the newly discovered HKU5-related *Merbecovirus* against 22 closely related HKU5 viruses.

**HKU5-r CoV spike glycoprotein characterization (BioProject PRJNA396502)**
The spike nucleotide sequence was translated to a protein sequence using ExPaSy (https://web.expasy.org/translate/) and a BLAST search was conducted against the NCBI nr database. The closest match was to the *Pipestrellus* bat coronavirus HKU5 with a 91.11% similarity. The high homology confirms its phylogenetic relationship with this clade (Supp. Fig. 5).

Multiple sequence alignment of the RBD of this novel sequence did not reveal any anomalies compared to other HKU5 sequences, however, one specific residue S566 was found to occur in MERS-CoV but not in HKU5 CoV or HKU4 CoV. The S1-S2 and S2' cleavage site sequences were found to be similar to other HKU5 sequences (Supp. Fig. 6).

**Human and camel DNA and other anomalies (BioProject PRJNA396502)**
Using BLAST analysis, we have obtained trace amounts of mitochondrial DNA from *Homo Sapiens* and *Camelus Dromedarius* from reads SRR5885860. We also obtained traces of *Pipistrellus Abramus* mitochondrial DNA, although the specific mitochondrial cytochrome oxidase I (COI) gene from the dataset turned out to be a perfect match to the reference genome which was isolated from South Korea, with significant differences from other field isolates of the COI gene from the same species (Supp. Fig. 7).

To help identify the source of the *Merbecovirus* sequences found in SRR5885860, we tested the other SRA run from the same experiment, SRR5885859 using NCBI STAT and fastv (Chen et al. 2020) and found that it did not contain any *Merbecovirus* sequences (Supp. Info 'Supplementary_Information_3_fastv_SRA_analysis.xls').

SRA's SRR5885151, SRR5885152 and SRR5885179 were then pooled together and bowtie2 was used to align the following genome sequences that were previously identified in the dataset using fastv: Rous sarcoma virus (NC_001407.1) with a genome read coverage of 72% (Supp. Fig. 8); *Mus musculus* mobilized endogenous polytropic provirus clone 15 truncated gag-pol polyprotein (gag) and envelope protein (env) genes, complete cds (NC_029853.1) with a genome read coverage of 59% (Supp. Fig. 9) and avian *myeloblastosis* virus RNA-dependent DNA polymerase gene, partial cds; transforming protein gene, complete cds; and long terminal repeat (NC_043404.1) with a genome read coverage of 93% (Supp. Fig. 10). All of which we were surprised to see in a cotton sequencing dataset.

**HKU5-related isolate BY140568 hits in BioProject PRJNA380842**
In the HZAU cotton sequencing BioProject PRJNA380842 uploaded to GenBank on September 28, 2018, we have identified HKU5-related CoV sequences in SRA run SRR7896912 using NCBI STAT (Supp. Fig. 11). Although we were unable to assemble a full-length genome due to the low amount of reads available, multiple reads returned 99% to 100% matches to the *Pipistrellus Abramus* bat coronavirus HKU5-related isolate BY140568, accession number MN611520.1 (Supp. Fig. 12). A total read coverage of the MN611520.1 genome was found to be 26% (Supp. Fig. 13).

The similarity of the HKU5-related *Merbecovirus* contig sequences in SRR7896912 to the most similar sequence(s) on the NCBI nr/nt database can be found in Supp. Table 1.

We queried the NCBI database using mitochondrion reference genome sequence accession NC_005436.1 as a query against experiment SRX4734343 as discussed in methods (Supp. Fig. 14). Surprisingly, instead of matches to the bat host *Pipistrellus Abramus*, BLAST hits returned mitochondrial DNA of *Sus Scrofa* (pig) and one read of *Camelus Dromedarius* (Supp. Table 2), despite the presence of a large number of reads that were identical to the *Pipistrellus abramus* bat coronavirus HKU5-related isolate BY140568 complete genome (accession MN611520.1) in the dataset.

A vector containing M13 Reverse primers and T7 and SP6 promoters was identified in de novo assembled contig k79_969 from SRR7896912 but of uncertain origin (Supp. Fig. 31).

**BSL-3 pathogen in cotton sequencing BioProject PRJNA380842**

Multiple non plant related viruses were identified in the *Gossypium hirsutum* (cotton) sequencing BioProject PRJNA380842 using the NCBI STAT tool. Fastv was also used for analysis of runs SRR7898912, SRR7896916, SRR7896926 and SRR7896936 from this project (Supp. Info. 'Supplementary_Information_3_fastv_SRA_analysis.xls'). Japanese encephalitis virus (JEV), a BSL-3 pathogen, was identified and we thus ran further BLAST analysis. The phylogenetic analysis result of SRR7896916, showing 16Kbp of sequences identified as "japanese encephalitis virus", and the distribution of the top 125 BLAST hits sequences for SRR7896916 to the subject MN639770.1 closest complete genome of JEV are shown in Supp. Fig. 15.

Due to the fragmented nature of the aligned genome, a total of 19 contigs were obtained from 121 reads using PRABI-Doua CAP3 (Huang and Madan, 1999) which all unambiguously map to wild isolates of the Japanese encephalitis virus (Supp Table 3). The alignment of raw reads in SRR7896916 to MN639770.1 is shown in Supp. Fig. 16.

In addition, the following genomes were identified in BioProject PRJNA380842: Wenzhou sobemo-like virus 4 strain mosZJ3539 (accession NC_033138.1) in dataset SRR7896926 and porcine circovirus 1 and Porcine circovirus 2 in dataset SRR7896916 (Supp. Info. Supplementary_Information_3_fastv_SRA_analysis.xls). Alignment of SRR7896926 to NC_033138.1 was conducted and shown in Supp. Fig. 17.

**HKU4-r CoV in silico infectious clone in SRR10915173 from BioProject PRJNA602160**

We *de novo* assembled SRR10915173 using MEGAHIT (Li et al. 2015). We then used NCBI BLASTn suite to align two sequences with BtTp-BetaCoV/GX2012 (accession KJ473822.1) as the query sequence against *de novo* assembled contigs as subject. We identified the longest matching contig with high homology to BtTp-BetaCoV/GX2012 (Supp. Fig. 18). From this contig we were able to directly assemble a complete genome sequence of a HKU4-related Merbecovirus (Fig. 8).

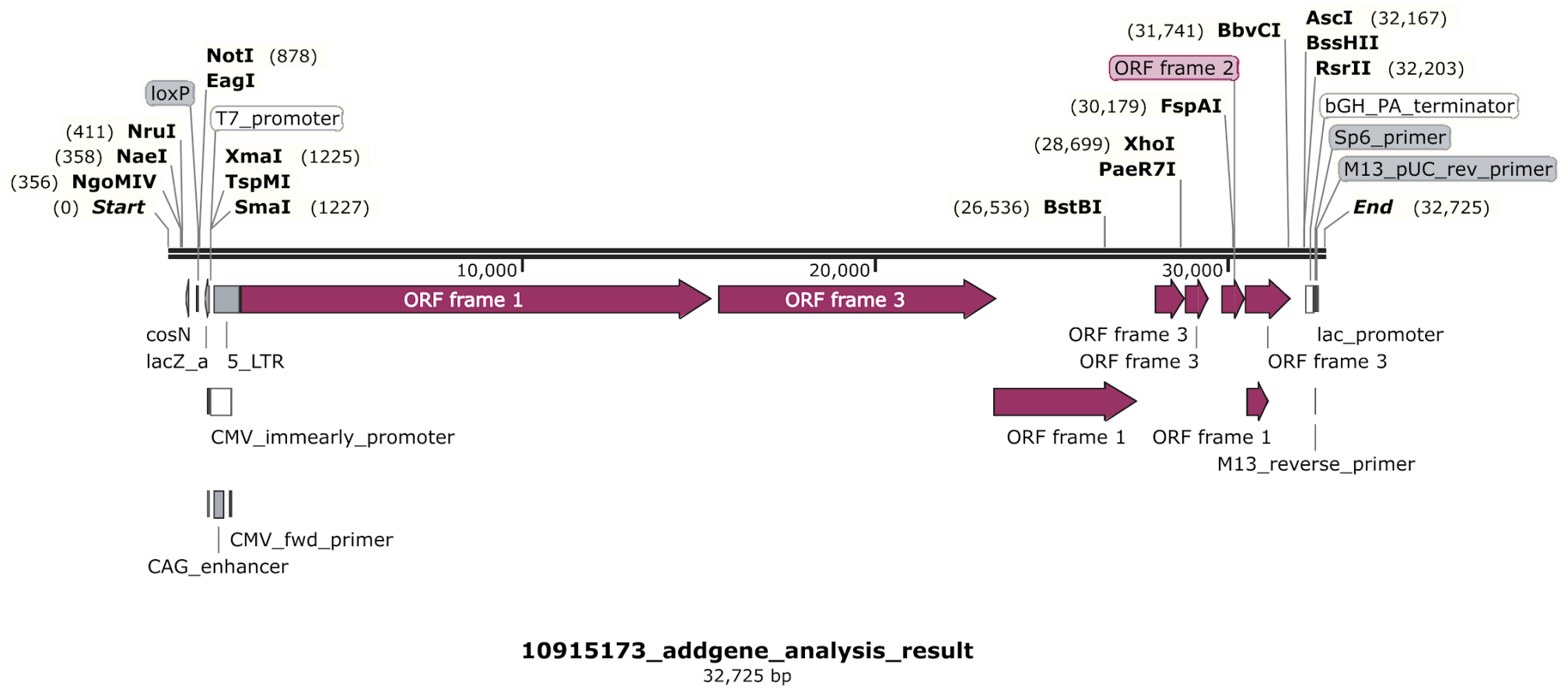

Fig. 8. Linear sequence map from the largest contig from MEGAHIT assembly with homology to BtTp-BetaCoV/GX2012 extracted from SRR10915173 displayed in SnapGene.

As the longest high homology contig was significantly longer than the length of the complete reference genome for BtTp-BetaCoV/GX2012 (32725 nt vs. 30247 nt), we attempted to obtain sequences that were found before the 5'-end and after the 3'-end of the novel coronavirus genome.

A BLAST analysis identified these sequences as being homologous to cloning vectors of diverse origin, with a CMV promoter before the 5'-end of the novel HKU4-related CoV genome and a Hepatitis D virus ribozyme followed by a bgH polyA signal after the 3'-end of the genome (Supp. Fig. 19). This indicates that the novel HKU4-related CoV obtained from SRR10915173 is likely an infectious clone as it is evidently in a format intended to generate full-length infectious RNA in mammalian cells.

To confirm our findings, we have identified the presence of the CMV promoter before the 5'-end and Hepatitis D virus ribozyme followed by bgH polyA signal after the 3'-end using coronaSPAdes and SPAdes *de novo* assembly contigs (Supp. Figs. 20, 21, 22, 23).

QC plots of raw read coverage and depth of the largest single contigs with homology to BtTp-BetaCoV/GX2012 from MEGAHIT assembly (Supp. Fig 24) and SPAdes assembly (Supp. Fig. 25) showed sufficient read coverage over the full HKU4-related clone sequence to have good confidence in the interpretation.

**HKU4-related CoV identified as a novel Merbecovirus (SRR10915173, BioProject PRJNA602160)**

A BLAST search of the HKU4-containing contig sequence against the NCBI nr/nt database revealed that it is only 98.38% similar to the closest known sequence of HKU4, BtTp-BetaCoV/GX2012 (Supp. Fig. 26).

A recombination analysis using RDP5 (Martin et al. 2020) did not reveal clear evidence that the purported infectious clone of a HKU4-related CoV we obtained from SRR10915173 was the result of simple recombination from any known sequences of HKU4 (Fig. 9).

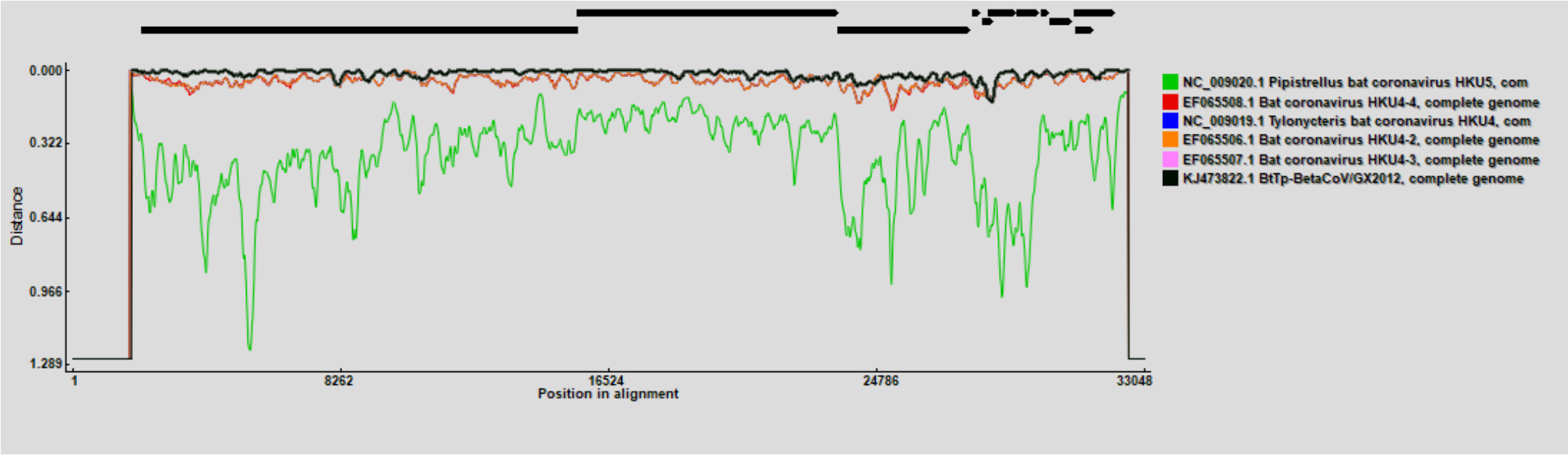

Fig. 9. RDP5 similarity plot of the novel HKU4-related CoV against other isolates of HKU4 and HKU5.

A table showing nucleotide similarity percentage of the newly discovered HKU4-related CoV purported infectious clone to other sequences on GenBank, based on BLAST search is shown in Table 1.

| ORF1ab | | | | S | | | |
|---|---|---|---|---|---|---|---|
| | SRR10915173 | KJ473822.1 | NC_009019.1 | | SRR10915173 | KJ473822.1 | NC_009019.1 |
| SRR10915173 | 100 | | | SRR10915173 | 100 | | |
| KJ473822.1 | 98.85 | 100 | | KJ473822.1 | 96.62 | 100 | |
| NC_009019.1 | 96.04 | 95.97 | 100 | NC_009019.1 | 92.86 | 92.78 | 100 |

| E | | | | M | | | | N | | | |
|---|---|---|---|---|---|---|---|---|---|---|---|
| | SRR10915173 | KJ473822.1 | NC_009019.1 | | SRR10915173 | KJ473822.1 | NC_009019.1 | | SRR10915173 | KJ473822.1 | NC_009019.1 |
| SRR10915173 | 100 | | | SRR10915173 | 100 | | | SRR10915173 | 100 | | |
| KJ473822.1 | 99.60 | 100 | | KJ473822.1 | 98.33 | 100 | | KJ473822.1 | 99.06 | 100 | |
| NC_009019.1 | 98.80 | 99.20 | 100 | NC_009019.1 | 96.21 | 96.52 | 100 | NC_009019.1 | 96.86 | 96.86 | 100 |

Table 1: Nucleotide similarity (in %) of the novel HKU4-related CoV purported infectious clone to other sequences on GenBank.

We then performed a BLAST search on the novel HKU4-related CoV clone against other known sequences of HKU4 on NCBI. This revealed that this sequence is only 98.38% similar to the closest known sequence of HKU4, BtTp-BetaCoV/GX2012, with critical components including the spike glycoprotein and the membrane glycoprotein bearing substantial differences from other known sequences of HKU4. This confirmed that this is a novel HKU4-related coronavirus that does not match any known published sequence. We therefore note that at least some unpublished viruses have been cloned and studied in laboratories in Wuhan.

In order to exclude the possibility that some of the sequences may have originated from a bat sample, we performed a BLAST search using the available nucleotide sequence of the bat *Tylonycteris pachypus*, the reservoir host of HKU4, against SRR10915173. We did not obtain any sequence that matches a known nucleotide sequence from this species.

**Novel HKU4-r CoV clone receptor binding domain may bind to human DPP4 (SRR10915173, BioProject PRJNA602160)**

Because the receptor-binding domain (RBD) of HKU4-related CoVs are known to bind human DPP4 (Wang et al. 2014), we undertook BLAST alignment of the RBD of the HKU4-related coronavirus clone to the closest matched sequence on PDB, 4QZV:B (Fig. 10). Then we performed modeling of the spike protein RBD of the novel HKU4-related coronavirus and its docking to human DPP4 using SWISS-MODEL and PRODIGY (Fig. 11).

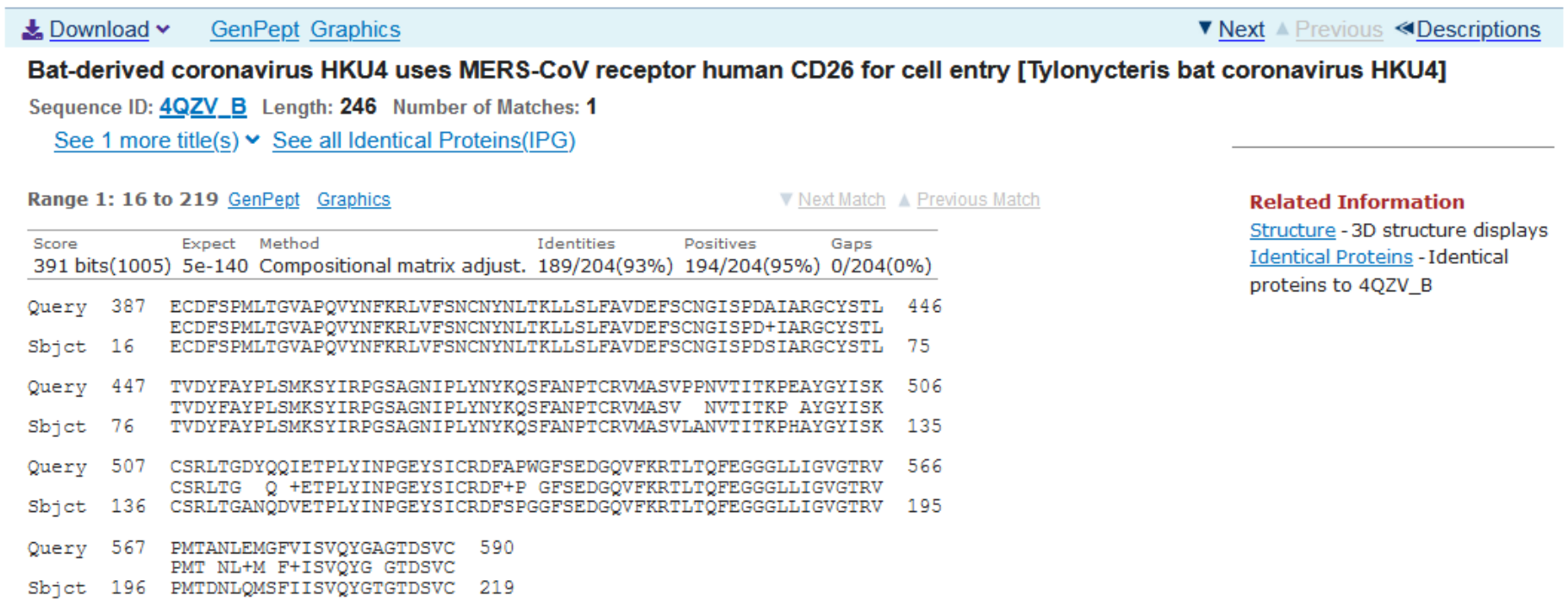

Download GenPept Graphics Next Previous Descriptions

**Bat-derived coronavirus HKU4 uses MERS-CoV receptor human CD26 for cell entry [Tylonycteris bat coronavirus HKU4]**

Sequence ID: 4QZV_B Length: 246 Number of Matches: 1

See 1 more title(s) See all Identical Proteins(IPG)

Range 1: 16 to 219 GenPept Graphics Next Match Previous Match

| Score | Expect | Method | Identities | Positives | Gaps |
|---|---|---|---|---|---|
| 391 bits(1005) | 5e-140 | Compositional matrix adjust. | 189/204(93%) | 194/204(95%) | 0/204(0%) |

```
Query  387   ECDFSPMLTGVAPQVYNFKRLVFSNCNYNLTKLLSLFAVDEFSCNGISPDAIARGCYSTL  446
             ECDFSPMLTGVAPQVYNFKRLVFSNCNYNLTKLLSLFAVDEFSCNGISPD+IARGCYSTL
Sbjct  16    ECDFSPMLTGVAPQVYNFKRLVFSNCNYNLTKLLSLFAVDEFSCNGISPDSIARGCYSTL  75

Query  447   TVDYFAYPLSMKSYIRPGSAGNIPLYNYKQSFANPTCRVMASVPPNVTITKPEAYGYISK  506
             TVDYFAYPLSMKSYIRPGSAGNIPLYNYKQSFANPTCRVMASV  NVTITKP AYGYISK
Sbjct  76    TVDYFAYPLSMKSYIRPGSAGNIPLYNYKQSFANPTCRVMASVLANVTITKPHAYGYISK  135

Query  507   CSRLTGDYQQIETPLYINPGEYSICRDFAPWGFSEDGQVFKRTLTQFEGGGLLIGVGTRV  566
             CSRLTG  Q +ETPLYINPGEYSICRDF+P GFSEDGQVFKRTLTQFEGGGLLIGVGTRV
Sbjct  136   CSRLTGANQDVETPLYINPGEYSICRDFSPGGFSEDGQVFKRTLTQFEGGGLLIGVGTRV  195

Query  567   PMTANLEMGFVISVQYGAGTDSVC  590
             PMT NL+M F+ISVQYG GTDSVC
Sbjct  196   PMTDNLQMSFIISVQYGTGTDSVC  219
```

Related Information

Structure - 3D structure displays

Identical Proteins - Identical proteins to 4QZV_B

Fig. 10. BLAST alignment of the RBD of the HKU4-related Coronavirus clone to the closest matched sequence on PDB, 4QZV:B.

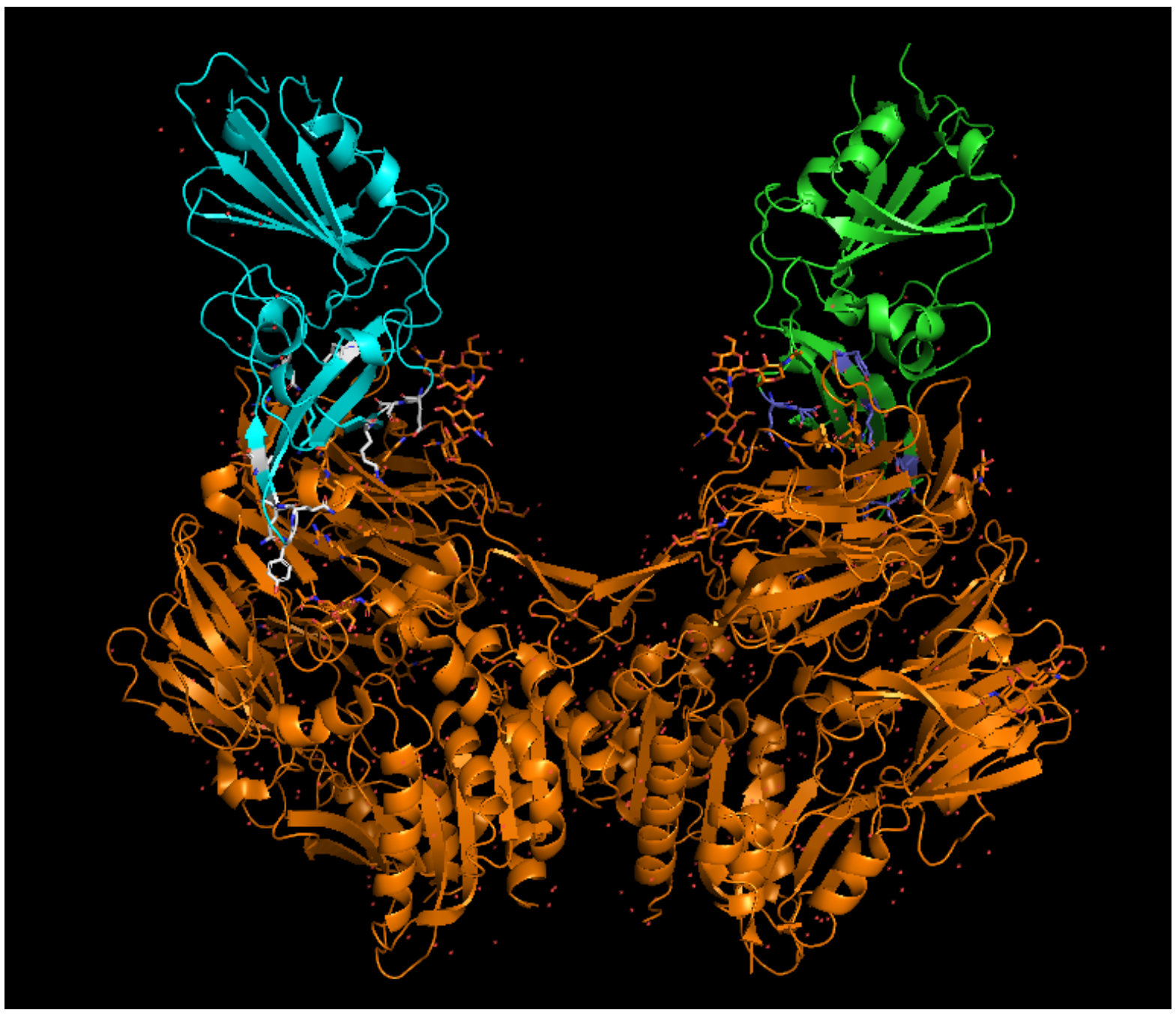

Fig. 11. Alignment of the modeled RBD of the HKU4-r clone (light blue) to the HKU4 RBD in PDB id: 4QZV (green) is consistent with binding to human DPP4 (orange). Sticks indicate residues contacting between the two molecules.

High structural homology between the RBD from the clone and the RBD of PDB id: 4QZV was observed (Fig. 11). Molecular docking modelling in PRODIGY showed comparable binding energy of the two RBD molecules to human DPP4 (Table 2), indicating that the purported infectious clone of HKU4 obtained from SRR10915173 is likely transmissible in human cells.

| RBD protein | Binding free energy (Kcal/Mol) | Predicted Kd(M) at 25℃ |
|---|---|---|
| HKU4 from Rice | -10.4 | 2.4e-8 |
| 4QZV:B | -10.5 | 2.0e-8 |

Table 2: PRODIGY binding energy and predicted binding affinity of the novel

HKU4-related Coronavirus Clone and Human DPP4 as compared to that of PDB:4QZV

**MERS-CoV contigs found in BioProject PRJNA602160**

During analysis of *de novo* assembled data from BioProject PRJNA602160, we also recovered four 239- to 318-nt long contigs in SRR10915173 and two contigs with lengths of 856 and 1011 nt from SRR10915174 with identical sequences to MERS-CoV.

**Novel HKU3-r CoV identified in a mouse sequencing dataset in BioProject PRJNA393936**

A significant amount of reads of a SARS-r CoV most closely related to HKU3 was identified in a *Mus Musculus* splenocyte RNA-seq sequencing dataset SRR5819071 in BioProject PRJNA393936. The novel bat SARS-like HKU3-related CoV was identified via BLAST analysis of the dataset, with the reads only 96-98% similar to their closest related GenBank sequences (Supp. Fig. 27, Supp. Table 4). Although the sequences are far too few to be assembled, the presence of transcription regulatory sequence (TRS)-leader associated fusion reads between the TRS-leader and the TRS-N gene suggest active replication of the HKU3-r CoV within the *Mus Musculus* splenocyte sample.

**SARS-WIV1 found in BioProject PRJNA358135**

In another rice sequencing project, but outside of the Hubei province, raw sequence reads of *Oryza sativa L.* from BioProject PRJNA358135 (Zhong et al., 2019) published to GenBank on September 1, 2017 by the Fujian Agriculture and Forestry University in Fouzou City, China, were also identified as having anomalous taxonomy using the NCBI STAT tool. Eight SRA datasets were assembled as per the methods section and the contigs were blasted against the BLAST nt database and numerous hits were found to bat SARS-r-CoV's and the SARS-like coronavirus WIV1 complete genome (KF367457.1) was identified as having the highest homology. The datasets were aligned to KF367457.1 and consensus sequences from two alignment methods both had 94.61% coverage of KF367457.1 with a combined 57 SNP difference to the reference genome (Supp. Fig. 28). Two primer sequences WIV1-NF "ATGTCTGATAATGGACCCCA" and MulPri-F7 "AGCCACCACATTTTCACCGAGGC", indicative of a possible synthetic construct, were identified in a de novo assembled contig, which was plotted in Addgene (Supp. Fig. 29) and was found to align with the SARS CoV WIV1 N protein sequence (Supp. Fig. 30).

## Discussion

Using a forensic bioinformatics approach we have identified numerous viruses anomalous to the sequencing of rice and cotton datasets from the HZAU and the FAFU, and a *Mus musculus* sequencing project from the WIV in China. We have found evidence of cross-contamination by multiple coronavirus strains indicating historical and ongoing research programs. In light of the current COVID-19 pandemic, we raise concerns about the handling and risk of release of potentially infectious human pathogens without urgent and significant improvements in biosafety protocols employed by these laboratories.

**Contamination Scenarios**

Cross-contamination of samples is a known issue in biological sequencing laboratories (Lusk, 2014; Selitsky et al. 2020). Cantalupo and Pipas (2019) - focussing on viral metagenomic studies - and Ballenghien et al. (2017) discuss the steps and contamination risks for NGS metagenomic sequencing projects, here summarized and extended with a focus on viral contamination: Sample collection ('passenger' virus contamination); Sample storage (sample degradation, microorganism growth (with incorrect temperature)); Shipping (cross contamination), Virus purification (sample loss); Nucleic acid purification (reagent contamination, possible loss of virus families depending on method, amplification bias); and Sequencing (machine contamination, index hopping, sequencing errors).

We do not believe that contamination by RNA plays a role in any of the findings we report here. RNA is fragile and any traces left in, for example, uncleaned vials or on surfaces would be quickly destroyed by Ribonuclease (RNase) activity ubiquitous in the environment. It is also unlikely that the same equipment would be used by two labs (WIV and HZAU) in two separate sittings without cleaning.

The most likely source for contamination for each project and in specific cases for each run is documented below. While we cannot be certain about all of the scenarios below, we proffer what we believe to be the most likely explanation in each case.

*Cotton sequencing BioProject PRJNA380842*
In relation to run SRR7896912, no bat material was found in this cotton sequencing BioProject, yet SRR7896912 was found to contain a significant amount of reads from an HKU5-related CoV. The source of this coronavirus is thus unlikely to have come from a bat sample. We hypothesize a possible explanation for this conundrum which should be further investigated: the final sequence loaded to GenBank MN611520.1 may have been assembled as DNA, and sequenced on the same lane as SRR7896912. The HKU5-related CoV contamination then occurred due to index hopping during sequencing. An actual bat sample was sequenced on a second lane. High quality reads from synthetic DNA were then filtered and added to the actual bat sample to form RNA sequencing study SRR11085738 ('RNA-Seq of Pipistrellus abramus: anal swab') from BioProject PRJNA606159 (Li et al. 2020).

In run SRR7896916, the significant amount of JEV reads, likely of pig cell culture sequencing origin, can most parsimoniously be explained by index hopping during sequencing of pig cell culture on the same lane as run SRR7896916.

*Rice sequencing BioProject PRJNA602160*
In relation to runs SRR10915173 and SRR10915174, the source of the novel HKU4 CoV purported infectious clone and MERS-CoV reads are unlikely to be sourced from index hopping due to the unique library format and protocol of the RNA-seq for this BioProject. RNA-seq of the cotton dataset was conducted with SMART-seq V4 Ultra low input RNA kit, sheared with AMPure beads and the resulting 200-400bp cDNA prepared with a TruSeq CHIP DNA kit. The RNA was isolated directly from zygotes collected with a micromanipulator system with FDA staining, the resulting single cells constructed individually with one library for each cell (https://www.ncbi.nlm.nih.gov/sra/?term=SRR10915173). The micromanipulator and staining before direct RNA extraction and input into low input cDNA kit, followed by a unique library construction, implies that the protocol was conducted in a specialized agricultural laboratory with no external laboratory work from the isolation through sequencing stages.

Furthermore, all WIV coronavirus sequencing projects on NCBI we have found to be paired end. The single end layout of BioProject PRJNA602160 is incompatible with paired end sequencing as they have entirely different chemistry cycle settings and cannot be multiplexed together.

It is exceedingly unlikely that a vial of MERS-CoV RNA would be brought into an agricultural laboratory and opened simultaneously with plant RNA isolation. Furthermore, if MERS-CoV RNA was extracted at the time of cross-contamination (e.g. on hands or on used pipettes) then ubiquitous RNAses would likely destroy the fragile RNA.

We believe that the most likely scenario of contamination of SRR10915173 and SRR10915174 by *Merbecovirus* reads (novel HKU4-r CoV and MERS-CoV) is by intact capsid cross-contamination via small quantities of aerosolized virus particles directly contracted from surfaces of potential media of cross-contamination prior to RNA isolation.

*Cotton sequencing BioProject PRJNA396502*
As novel HKU5-r CoV reads were found in run SRR5885860 but not run SRR5885859, which was part of the same experiment, we believe this indicates that index hopping during sequencing is the most likely cause of contamination.

*Rice sequencing BioProject PRJNA358135*
SARS-WIV1 CoV is only found in the runs in the project with names including the text 'GY11'. However, both japonica rice cultivar Nipponbare ('JPN') and an indica rice cultivar 93-11 ('9311') datasets contain SARS-WIV1 (Supplementary_Information_2_SRA_datasets.xlsx). Moloney Murine Leukemia virus is found in 13 runs, but none of these runs contain SARS-WIV1. We infer that the bat CoV SARS-WIV1 and Moloney Murine Virus material is likely to have entered the rice dataset during the preparation stage prior to RNA extraction/library construction. Potentially during collection or through handling of the GY11 sample rack by humans with bat cell media on hands/gloves.

*Mus musculus sequencing BioProject PRJNA393936*
Novel HKU3-r CoV reads were found only in run SRR5819071 of the 4 *Mus Musculus* datasets. We infer the most likely source of contamination is index hopping from a related bat sequencing project.

**Cross-contamination between different laboratories in Wuhan and BSL-3 breaches**
The discovery of a novel HKU4-r CoV purported infectious clone, and of several contig sequences identical to MERS-CoV (a BSL-3 pathogen) in an agricultural sample of *Oryza sativa Japonica* sequenced by HZAU, indicates that sequences from multiple organisms of differential biosafety levels were circulating within the laboratory systems of Wuhan, and crossing institutional boundaries from the BSL-3 laboratory of WIV to a non-pathogen-related agricultural laboratory of the HZAU. The discovery of both cloned HKU-4 and fragments of MERS-CoV in a dataset published to GenBank on February 9, 2020, along with the discovery of an unexpected BSL-4 organism (Nipah Henipahvirus) co-occurring with a BSL-2 organism (Spodoptera frugiperda rhabdovirus) in the same SRA as the very first sequences of SARS-CoV-2 (Zhang, 2021) indicates that BSL-3 laboratories had already been breached at WIV at the beginning of the SARS-CoV-2 pandemic.

The discovery of HKU5-related isolate BY14068 reads in the cotton-targeted deep sequencing project PRJNA380842 by the HZAU is also concerning. This discovery indicates a direct cross-contamination event between the WIV and the HZAU, at a facility likely designed for sequencing low-risk agricultural samples. Furthermore, SRR7896912 sequencing was published on September 28, 2018, while MN611520.1 was published to GenBank over a year later on October 24, 2019 by the WIV. As we are unable to find sequences in the dataset related to *Pipistrellus spp.,* the bat host of HKU5, this also raises serious questions on the legitimacy of these and potentially other sequences, BioSamples and SRA runs deposited to GenBank by the WIV.

Also in the HZAU BioProject PRJNA380842, the discovery of a BSL-3 pathogen JEV in SRR7896916 is troubling. The biological agent of study (*Gossypium hirsutum*) is a low risk agent with no potential to cause disease or economic loss. Further, the samples were prepared and sequenced in an agricultural laboratory without the requirement for the facility, safety equipment and protocols of a BSL-3 pathogen biosafety laboratory for which isolates of the high-risk human pathogen JEV are required to be handled. We note the publication of an experiment using JEV within the HZAU using pig (*Sus Scrofa*) cells (Zheng, 2019). As the Mitochondrial sequences of *Sus Scrofa* have been found in SRR7896916 (Supp. Table 3), we infer the likely origin of the JEV reads from SRR7896916 to be a pig cell infection experiment. As all infection experiments using wild-type (field isolate) of JEV require BSL-3 conditions, the escape of this JEV sample and its subsequent introduction into SRR7896916 constitutes a BSL-3 breach with the release of BSL-3 microorganisms within the HZAU facility. As the JEV strain MN639770.1 is a field (wild-type) strain originally isolated from human sera in Yunnan, with all its closest sequences in NCBI GenBank having origins in either Yunnan or Guangdong (but not Wuhan), the sample found within Wuhan is likely a known sequence previously isolated in culture. This hypothesis is strengthened by the observation that *de novo* assembled contigs were found to be of the same level of similarity (same E value in BLAST result) to both MN639770.1 and to MH258848.1 (strain SA14), both of which are non-attenuated non-vaccine strains requiring BSL-3 conditions for analysis.

The occurrence of mitochondrial DNA from *Homo Sapiens* and *Camelus Dromedarius* in a *Gossypium barbadense* cotton leaf sequencing BioProject PRJNA396502 run SRR5885860 is highly surprising. Furthermore, trace reads of *Pipistrellus abramus* mitochondrial DNA isolated from South Korea likely indicates that these sequences were previously isolated in culture. As neither camels nor bats are within the scope of sampling for cotton leaves, and the other SRA run from the same experiment, SRR5885859, did not contain any *Merbecovirus* sequences, the origin of these sequences from within SRR5885860 is likely cross-contamination from within the sample-generating laboratory.

As SRR5885860 was sequenced and uploaded to GenBank by the HZAU, we searched PubMed for *Merbecovirus*-related research conducted by the HZAU in 2017 and 2018. We found only a single citation (Chen et al. 2017), which used MERS-CoV (EMC/2012 strain), but did not use either *Pipistrellus abramus* or *Camelus dromedarius* material in their study.

Due to the lack of evidence of studies using the bat genus *Pipistrellus* or camelids at the HZAU at the time of SRR5885860 creation, we infer that the *Merbecovirus*, camel and bat sequences within SRR5885860 represent a cross-contamination event by a virology-related laboratory outside of the HZAU. As MERS-CoV is considered a BSL-3 organism, and MERS-r CoVs are considered BSL-2 organisms, the presence of a MERS-r CoV sequence appears to represent an escape of the MERS-r CoV material from a laboratory with a biosafety level of BSL-2 at a minimum, and its subsequent introduction into a different laboratory. We infer the WIV to be a possible source, given their history of MERS-CoV and *Merbecovirus* research (Luo et al. 2018, Yang et al. 2015), especially considering that sequencing efforts were mostly outsourced by WIV to other laboratories in Wuhan. In an interview by Science Magazine in 2020 Zhengli Shi from WIV was asked “Does your group extract viruses from biological samples and do the sequencing or does that take place elsewhere?”. To which Zhengli Shi responded “We isolated viruses or extracted virus RNA from biological samples in the lab. The sequencing was done mostly in Wuhan.” (https://www.sciencemag.org/sites/default/files/Shi%20Zhengli%20Q&A.pdf).

The identification of a SARS-r CoV most closely related to HKU3 in a *Mus Musculus Splenocyte* RNA-seq dataset SRR5819071 by the WIV confirms that significant cross-contamination also occurs between different samples within the WIV.

**Viral sequences unpublished by the WIV**

The WIV is known to keep multiple undocumented viral sequences (Sirotkin and Sirotkin, 2020). Here we document for the first time two unpublished complete viral sequences. We have isolated and assembled the full-length sequence of a novel HKU5-related CoV from a HZAU cotton sequencing dataset accession SRR5885860 (BioProject PRJNA396502). The dataset was uploaded to GenBank on December 16, 2017. The novel HKU5-related CoV is not closely related to any known sequence of HKU5 currently available in the NCBI nucleotide database, and to this date has not been published by the WIV. We note that of the five most closely related viruses, two were found in South Korea (Lee et al. 2018), while for the 3 found in China, accession DQ648802 was collected by Tang et al (2006), the other two (accessions KX285199.1 and KX285200.1) were collected during a visit to Nanling (Guangdong) made by WIV on June 18, 2013 (https://www.ncbi.nlm.nih.gov/nuccore/KX285199.1, https://www.ncbi.nlm.nih.gov/nuccore/KX285200.1). These two samples also belong to PREDICT (Anthony et al. 2017) and were published to GenBank on June 12, 2017, although their sequencing was done using “traditional Sanger dideoxy sequencing”.

We also document the discovery of a novel HKU4-r CoV in a plant (Oryza Sativa Japonica) sequencing project (PRJNA602160) from an agricultural laboratory from the HZAU. Furthermore, we have discovered that the virus has been cloned with the same vector format as a clone of Nipah Henipahvirus from BioProject PRJNA605983 (Zhang, 2021), a dataset which contained bronchoalveolar lavage fluid samples collected by the Wuhan Jinyintan Hospital on December 30, 2019 from seven patients infected with SARS-CoV-2 and sequenced by WIV (Zhou et al. 2020). The BioProject for the Oriza Sativa Japonica sequencing was registered at GenBank on January 19, 2020, but to this date the HKU4-r CoV has not been published by the WIV. We note that *Oryza sativa* cultivar:japonica (rice) sequencing BioProject PRJNA601977 was registered on NCBI on the 17/01/2020 by the WIV, 2 days before registration of the *Oryza sativa* Japonica BioProject PRJNA602160 by the HZAU (containing the novel HKU4-r CoV clone), indicating the two projects may be related, however this observation cannot be confirmed as no data has been published for PRJNA601977 by the WIV.

We further document the discovery of an unpublished novel HKU3-r CoV in *Mus musculus* sequencing run SRR5819071 in WIV BioProject PRJNA393936 published to GenBank on the July 12, 2017, although the read quantity is not sufficient to assemble a complete genome.

Regarding the potential connection of our findings to uncovering SARS-CoV-2 origins, one of the main arguments for the hypothesis that SARS-CoV-2 could not be have been an engineered virus, "if genetic manipulation had been performed, one of the several reverse-genetic systems available for *Betacoronaviruses* would probably have been used" (Andersen et al. 2020) is clearly invalidated by our work. There are no known publications documenting a reverse genetic system on the strain BtTp-BetaCoV/GX2012, from which the backbone of the novel HKU4-r CoV purported human infectious clone identified here was derived. Compared with BtTp-betaCoV-GX2012 we note minimal insertions: notably the insertion of the sequence TTT between A5555-A5558 and the sequence TGTTCC between A8183-T8190 in ORF1a. This is quite unexpected for directed engineering, indicating that an unpublished backbone was used. Neither the exact sequence nor the reverse genetic system that was used to construct this sequence has been published.

The fact that this infectious clone was discovered in an unlikely dataset, as a contaminant in a HZAU rice sequence bioproject, raises the question as to how many more potentially pathogenic coronaviruses have been researched by the WIV without disclosure to the international scientific community.

**Synthetically engineered construct cannot be identified as such from the HKU4-r CoV sequence alone**

In an attempt to find an obvious signature of genetic engineering in the sequence of this HKU4-r CoV purported infectious clone, we performed a restriction enzyme mapping of the sequence using SnapGene Viewer (SnapGene® software) using the set of all type II restriction endonucleases, with the addition of the enzyme BsmBI and BsaI, which are often used in conjunction for the construction of coronavirus reverse genetic systems (Fig 12). For comparison, we also obtained and performed restriction site mapping of two related coronaviruses: BtTp-BetaCoV/GX2012 (accession KJ473822.1) (Fig. 13) and HKU4-1 (accession NC_009019.1) (Fig. 14).

We did not observe a clear pattern of restriction site mapping between these sequences. Consequently, the mapping pattern for both type II and type IIS restriction endonuclease recognition sites across these sequences was not conserved. Querying of the sites that correspond to type II or type IIS restriction sites in these sequences did not reveal any universally conserved nucleotides across known sequences of HKU4 and MERS-CoV within NCBI databases, which is evident by the lack of conservation of these sites across these sequences.

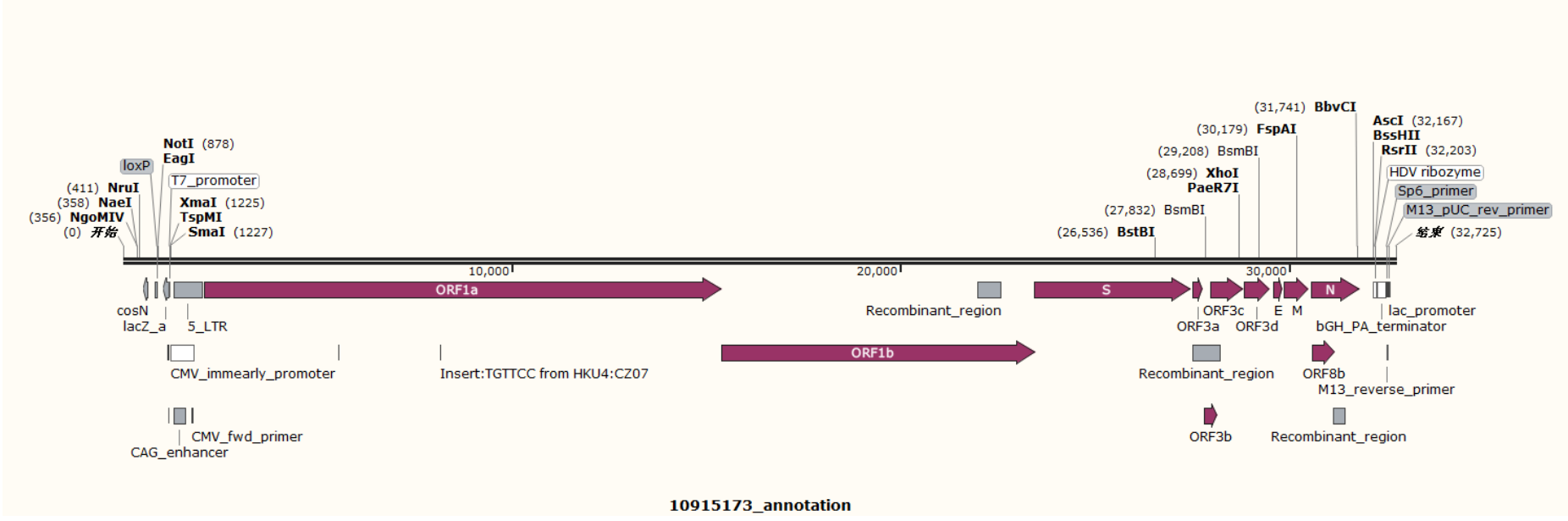

Fig. 12. Annotation and restriction mapping of the largest contig corresponding to the infectious clone of HKU4 from *de novo* assembly of SRR10915173

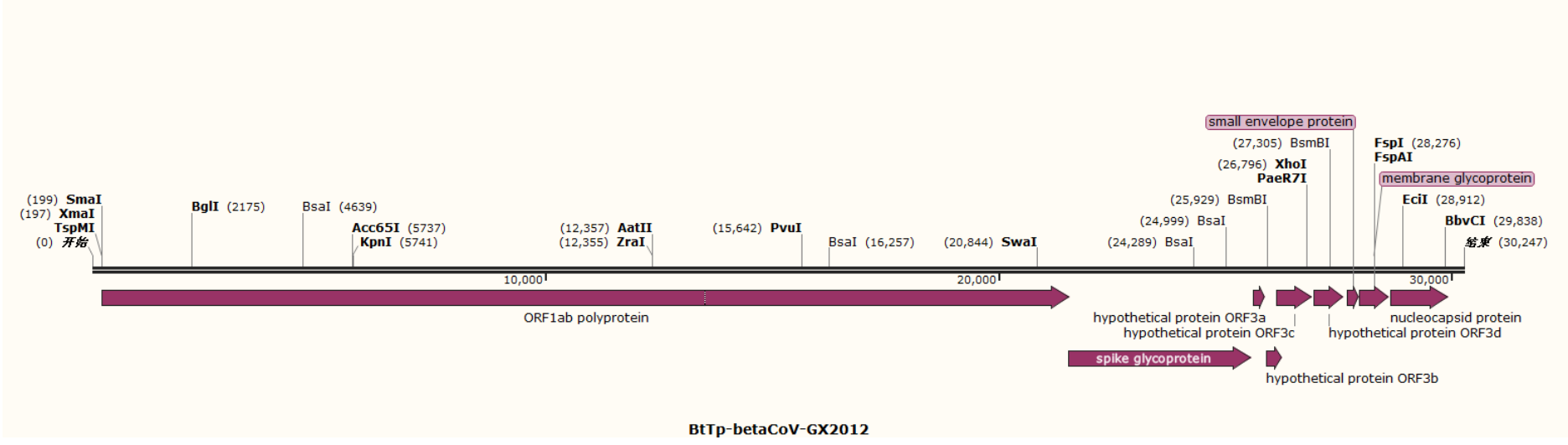

Fig. 13. Annotation and restriction mapping of BtTp-BetaCoV/GX2012 for comparison

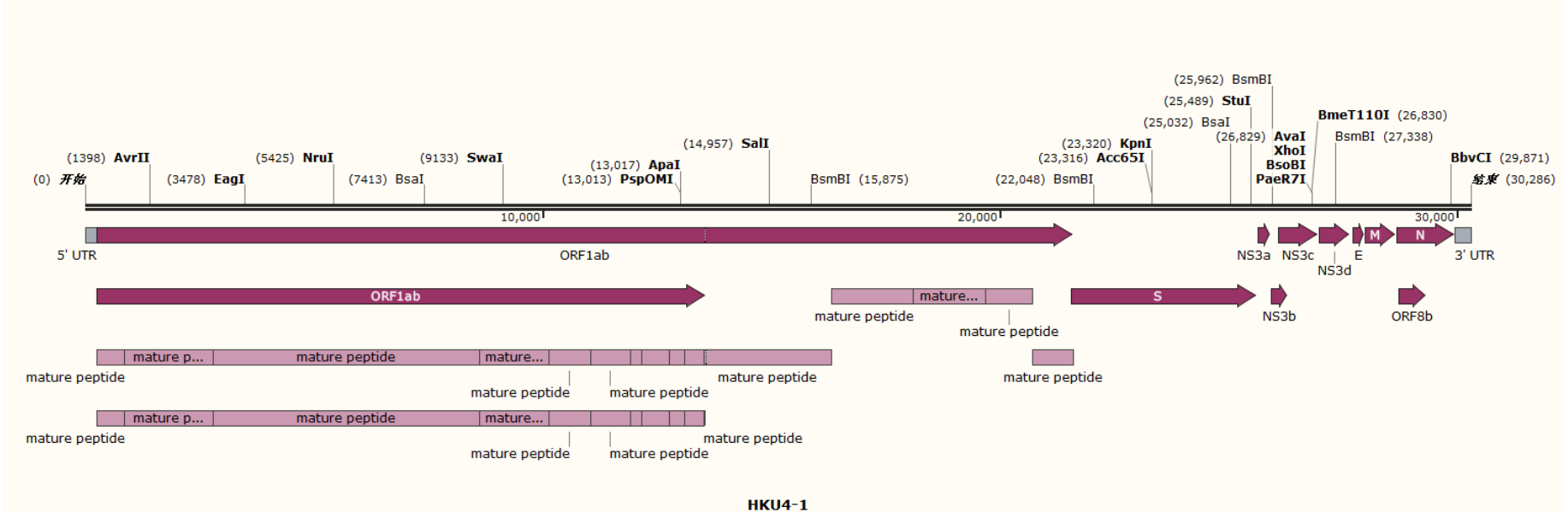

Fig. 14. Annotation and restriction mapping of HKU4-1 for comparison

In an attempt to obtain a potential previous reverse genetics system for HKU4-like coronaviruses, we also performed a keyword search using "HKU4" and "reverse genetics" or "infectious clone" using PubMed and Google Scholar. However, we found no scientific publications which disclose such a reverse genetic system, or have demonstrated the construction of an infectious clone for HKU4-related coronaviruses. We therefore conclude that no obvious signature of synthetic engineering with known reverse genetic systems could be deduced from any of the three sequences analyzed alone. Indeed, all 3 sequences have similar levels of cloning difficulty, and could all be cloned without inducing nucleotide changes as per when a newly disclosed reverse genetic system was developed recently to clone SARS-CoV-2 (Xie et al. 2021), or the very first type IIS restriction endonuclease based gene assembly system was used to clone the first known coronavirus (Yount et al. 2002).

The lack of obvious patterns for restriction sites in the newly discovered HKU4 clone and other wild isolates of HKU4 and MERS-CoV demonstrates the weakness of attempting to use assumed predictable genetic signatures left by detection for signatures of genetic laboratory engineering using properties of the viral sequences alone. This is especially true when the original sequence has been kept unpublished, making the closest related natural sequences inadequate for use as reference sequences due to the large evolutionary distances masking potential signatures within a large number of potentially natural changes during evolution from one sequence to another.

The existence of such a clone from the contamination of an SRA dataset sequenced and submitted by a laboratory in Wuhan, also demonstrates that systems which can be used to seamlessly assemble an infectious coronavirus without leaving any genomic fingerprints are in use by laboratories in Wuhan.

## Conclusion

In this paper, we have reported an unexpected discovery of several highly pathogenic viruses in the raw sequencing data from laboratories unsuited for animal and human pathogens. In particular, we report the identification and full-sequence assembly of a novel HKU5-related *Merbecovirus.* Additionally, we report the discovery of a novel HKU4-related *in silico* human infectious clone CoV in the same vector format as an infectious clone of *Nipah Henipah* virus which was found in the WIV metagenomic sequencing data of first SARS-CoV-2 infected patients (Zhang, 2021). We further report the discovery of a HKU5-r CoV in a cotton sequencing project by HZAU, over a year prior to its publication by WIV. Also, we have identified MERS-CoV sequences and JEV sequences in an agricultural dataset from HZAU and a nearly complete SARS-WIV1 assembly from a rice sequencing project by FAFU. The identification of viral sequences and clones originally sourced from the WIV in agricultural laboratories should be followed up by an independent forensic examination of the laboratories involved.

In addition, our characterization of infectious clone sequences has revealed that not only does WIV possess samples of unpublished viruses, they are also actively manipulating such viral sequences via the construction of infectious clones and potential genetic engineering of such clones, as discovered here, without disclosing the backbones used to construct such clones to the international scientific community.

In addition, the significant genetic distance from the newly discovered infectious clone of a HKU4-related coronavirus to other known sequences of HKU4 has resulted in the masking of potential signatures of genetic manipulation. The inability to detect significant signatures of genetic manipulation using sequences of this infectious clone and several related sequences of HKU4 illustrates the weakness of such methods against constructs generated using the type IIS enzyme-based "No See'm" technology used in current reverse genetic systems (Yount et al. 2002, Andersen et al. 2020).

Our discoveries provide evidence of a significant level of secrecy, as well as the presence of unpublished viral sequences and clones within various laboratories in Wuhan, including the Wuhan Institute of Virology. This demonstrates the need for a thorough forensic investigation into the various laboratories in Wuhan, especially into unpublished and non-viral short-read sequencing datasets that originated from such laboratories, in the ongoing investigation into the origins of SARS-CoV-2.

Due to our discovery of potential cross-contamination between biological samples in China, we also advise caution when using SRA data from China as evidence of virus-host connections when identifying and assembling viral sequences from metagenomic sequencing data. We further recommend a systematic data mining and virus discovery program on past SRA datasets from China, particularly from plants and other agricultural samples, which may reveal novel and unpublished viral sequences from cross-contamination between published and unpublished samples, as we have done here.

The importance of our work is threefold. The documentation of previously unidentified lax safety standards and biosecurity breaches between WIV and HZAU should provide an impetus to improve biosafety control in Wuhan and the greater People's Republic of China. Secondly, our discovery of an undocumented genetically modified *Merbecovirus* which cannot be identified as synthetically manipulated based on the HKU4-related CoV genetic sequence alone is a potentially limiting factor in future genetic engineering attribution for biosecurity safeguarding (Lewis et al. 2020). Finally, through our identification and publishing of the sequences of a novel HKU4-related CoV and a novel HKU5-related CoV, we enable the wider scientific community to carry out further genetic and phylogenetic studies on these strains.

## Acknowledgements

We acknowledge Francisco de Asis for providing the initial discovery and example sequence reads from HKU5-related Coronaviruses in SRR5885860 and SRR7896912, investigating on the usage history of Illumina sequencers by the WIV and the HZAU, and for helping with the phylogenetic tree analysis and discovery of additional evidences of cross-contamination within the NCBI SRA datasets that were deposited by institutes in Wuhan.

We also acknowledge Dan Sirotkin for review and feedback which helped improve the manuscript.

## Competing Interests

The authors declare that they have no competing interests.

## Data availability

**Supplementary Information**

Supplementary figures, tables and spreadsheets have been deposited on Zenodo.org under the doi: 10.5281/zenodo.4661667

Supplementary_Information_1_Figures_and_Tables.docx: Supplementary figures, tables and contigs
Supplementary_Information_2_SRA_datasets.xlsx: Summary of all SRA datasets used for this analysis, including sha256sum checksums (sha256sum (GNU coreutils) 8.30) for NBCI source datasets
Supplementary_Information_3_fastv_SRA_analysis.xlsx: Fastv viral analysis of each SRA used in this analysis.
Supplementary_Information_4_HKU5-1-Cotton-HZAU-SRR5885860_Blast_Tree_View.pdf: Pdf version of Figure 4.
Selected supplementary figures from Supplementary_Information_1_Figures_and_Tables.docx have been saved in .png format
Fig_9_novel_HKU4r_CoV_RDP5_distance_plot.bmp: Figure 9 has been saved in

.bmp format

**PRJNA602160**
Data generated via assembly and analysis of PRJNA602160 has been deposited on Zenodo.org under the doi: 10.5281/zenodo.4620604

All contig sequences with homology to KJ473822.1 found within PRJNA602160 can be found as 10915167.fa , 10915172.fa , 10915173.fa and 10915174.fa

Reads supporting the vector-virus junctions within SRR10915173 have been included in 10915173.fa

The raw MEGAHIT contig sequences from PRJNA602160 can be found as SRR10915167_final.contigs.fa.gz , SRR10915168_final.contigs.fa.gz , SRR10915173_final.contigs.fa.gz and SRR10915174_final.contigs.fa.gz

Sequences that were found to be near-identical to MERS-CoV can be found as MERS_CoV from SRR10915173.fa and MERS_CoV from SRR10915174.fa

The Addgene analysis results supporting fig.1 can be found as 10915173_addgene_analysis_result.gb

The annotated genome of the HKU4-related Coronavirus clone found in SRR10915173 can be found as  10915173_annotation.gb

Additional analysis of SRR10915173 were performed using CoronaSPADES, and can be found as  SRR10915173_coronaspades_default.tar.gz

The 3-dimensional model of the RBD of the HKU4-related Coronavirus Clone can be found as  HKU4_RBD.pdb

Sequences found with homology with the bat Tylonycteris pachypus, their similarity(identity/length of match) to the bat sequence, the most similar sequence on nt and their similarity(identity/length of match) to such sequences can be found as Bat_candidate.fa

Contig sequences that were identical to MERS-CoV in accession PRJNA602160 can be found as MERS_CoV from SRR10915173.fa and MERS_CoV from SRR10915174.fa.

SPAdes *de novo* assembly of SRR10915173 has been deposited on Zenodo.org under the doi: 10.5281/zenodo.4661667

**PRJNA396502**
Data generated via assembly of SRR5885860 from BioProject PRJNA396502 has been deposited on Zenodo.org under the doi: 10.5281/zenodo.4399248

HKU5-1-Cotton-HZAU-SRR5885860.fa:  Consensus sequence obtained through de novo assembly of SRR5885860 using EGassembler.

HKU5-1-Cotton-HZAU-SRR5885860.gb: Consensus sequence obtained through de novo assembly of SRR5885860 using EGassembler, annotated in Snapgene.

HKU5-2-Cotton_HZAU-SRR5885860.fa: Single config obtained through de novo assembly of SRR5885860 using MEGAHIT.

HKU5-2-Cotton_HZAU-SRR5885860.gb: Single config obtained through de novo assembly of SRR5885860 using MEGAHIT, annotated in Snapgene.

multalin.fa: EGassembler and MEGAHIT de novo assemblies of HKU5-r CoV aligned using MultAlin, in fasta format.

multalin.gif: EGassembler and MEGAHIT de novo assemblies of HKU5-r CoV aligned using MultAlin.

Bowtie2 alignment of pooled SRR5885851, SRR5885852, SRR5885879 to selected reference genomes has been deposited on Zenodo.org under the doi: 10.5281/zenodo.4661667

**PRJNA358135**
Data generated via alignment of 8 SRA's (SRR5127175, SRR5127184, SRR5127186, SRR5127187, SRR5127189, SRR5127191, SRR5127192, SRR5127197) from BioProject PRJNA58135 has been deposited on Zenodo.org under the doi: 10.5281/zenodo.4661667

SRP095383_8runs_KF367457_1_bowtie2_gatk_sorted_marked.bam: bowtie2 alignment of raw reads to KF367457.1
SRP095383_8runs_KF367457_1_bwa_mem_gatk_sorted_marked_extended.bam: bwa alignment of raw reads to KF367457.1
SRP095383_8runs_KF367457_1_bwa_mem_gatk_sorted_marked_igv_consensus.fa: consensus fasta file from bwa alignment, generated in IGV.
SRP095383_8runs_KF367457_1_bowtie2_gatk_sorted_marked_igv_consensus.fa: consensus fasta file from bowtie2 alignment, generated in IGV.

**PRJNA380842**
Data generated via alignments of SRR7896912, SRR7896916, SRR7896926 to selected reference genomes has been deposited on Zenodo.org under the doi: 10.5281/zenodo.4661667

SRR7896912_MN611520_1_bowtie2_sorted_marked.bam: Alignment of SRR7896912 raw reads to MN611520.1 using bowtie2.